\documentclass[12pt]{article}
\usepackage{amssymb}
\usepackage{amsmath}
\usepackage{epsfig}
\usepackage{graphicx}
\usepackage{wrapfig}

\begin{document}
\newcommand{\rrtth}{$\gamma\gamma \to t \bar t h^0$ }
\newcommand{\rrtthg}{$\gamma\gamma \to t \bar t h^0+g$ }
\newcommand{\eetth}{$e^+ e^- \to \gamma\gamma \to t \bar t h^0$ }
\newcommand{\eetotth}{$e^+ e^-  \to t \bar t h^0$ }

\title{ The effects of the little Higgs models on $t\bar{t} h^0$
production via $\gamma \gamma$ collision at linear colliders
\footnote{Supported by National Natural Science Foundation of
China.}} \vspace{3mm}

\author{{ Pan Kai, Zhang Ren-You, Ma Wen-Gan, Sun Hao, Han Liang, and Jiang Yi }\\
{\small  Department of Modern Physics, University of Science and Technology}\\
{\small  of China (USTC), Hefei, Anhui 230026, P.R.China} }

\date{}
\maketitle \vskip 12mm

\begin{abstract}
In the frameworks of the littlest Higgs($LH$) model and its
extension with T-parity($LHT$), we studied the associated $t\bar
th^0$ production process \eetth at the future $e^+e^-$ linear
colliders up to QCD next-to-leading order. We present the regions
of $\sqrt{s}-f$ parameter space in which the $LH$ and $LHT$
effects can and cannot be discovered with the criteria assumed in
this paper. The production rates of process \rrtth in different
photon polarization collision modes are also discussed. We
conclude that one could observe the effects contributed by the
$LH$ or $LHT$ model on the cross section for the process \eetth in
a reasonable parameter space, or might put more stringent
constraints on the $LH$/$LHT$ parameters in the future experiments
at linear colliders.

\end{abstract}

\vskip 5cm {\large\bf PACS: 12.60.Cn, 14.80.Cp, 14.65.Ha}

\renewcommand{\theequation}{\arabic{section}.\arabic{equation}}
\renewcommand{\thesection}{\Roman{section}}
\newcommand{\nb}{\nonumber}

\makeatletter      
\@addtoreset{equation}{section}
\makeatother       

\par
\section{Introduction}
\par
The standard model($SM$)\cite{s1}\cite{s2} of elementary particle
physics provides a remarkably successful description of high
energy physics phenomena at the energy scale up to $100~GeV$.
Despite its tremendous success, the mechanism of electroweak
symmetry breaking ($EWSB$) remains the most prominent mystery in
current particle physics, and the Higgs boson mass suffers from an
instability under radiative corrections leading to the "hierarchy
problem" between the electroweak scale and the $10~TeV$ cut-off
scale $\Lambda$. The study of the $EWSB$ and the "hierarchy
problem" motivate many research works on the extensions of the
$SM$. Recently, the little Higgs models have drawn a lot of
interests as they offer an alternative approach to solve the
"hierarchy problem"\cite{Arkani}, and were proposed as one kind of
models of $EWSB$ without fine-tuning in which the Higgs boson is
naturally light as a result of non-linearly realized
symmetry\cite{LH1}-\cite{LH7}. The most economical model of them
is the littlest Higgs($LH$) model, which is based on an
$SU(5)/SO(5)$ nonlinear sigma model\cite{LH5}. The key feature of
this kind of models is that the Higgs boson is a pseudo-Goldstone
boson of a global symmetry, which is spontaneously broken at some
higher scale $f$, and thus is naturally light. The $EWSB$ is
induced by a Coleman-Weinberg potential, which is generated by
integrating out the heavy degrees of freedom.

\par
In the $LH$ model without T-parity, a set of new heavy gauge
bosons ($A_H,Z_H,W_H$) and a new heavy vector-like quark ($T$) are
introduced which just cancel the quadratic divergences of Higgs
self-energy induced by $SM$ gauge boson loops and the top quark
loop, respectively. However, it has been shown that the $LH$ model
without T-parity suffers from severe constraints from the
precision electroweak data, which would require raising the masses
of new particles to be much higher than
$1~TeV$\cite{LHConstraints}. To avoid this problem, T-parity is
introduced into the $LH$ model, which is called the littlest Higgs
model with T-parity ($LHT$)\cite{LHT}. In the $LHT$ model, the
$SM$ particles are T-even and the most of the new heavy particles
are T-odd. Thus, the $SM$ gauge bosons cannot mix with the new
gauge bosons, and the electroweak precision observables are not
modified at tree level. Beyond the tree level, small radiative
corrections induced by the model to precision data still allow the
symmetry breaking scale $f$ to be significantly lower than
$1~TeV$\cite{Hub}. In the top-quark sector, the $LHT$ model
contains a T-odd and a T-even partner of the top quark. The T-even
partner of the top quark mixes with top quark and cancels the
quadratic divergence from top quark loop in the contributions to
Higgs boson mass. Consequently, the $LHT$ model could induce
abundant new phenomenology in present and future experiments.

\par
In previous works, it is concluded that the LHC has great
potential to discover directly the new particles predicted by the
little Higgs models up to multi-TeV mass scale, such as the
colored vector-like quark $T$, heavy gauge bosons and so on, in
Refs.\cite{Han}\cite{Belyaev} and the references therein. After
the new particles or interactions in the little Higgs models had
been directly discovered at the LHC experiment, the International
Linear Collider(ILC) would then play an important role in the
detailed study of these new phenomena and accurate measurement of
the interactions in the little Higgs models.

\par
The precise measurement of the process of $t\bar th^0$ production
at the ILC is particularly important for probing the Yukawa
coupling between top-quarks and the Higgs boson with intermediate
mass. Actually, the $t \bar{t} h^0$ production can be first
detected at the CERN LHC and further precisely measured at the
ILC. It was pointed out that the $t-\bar t-h^0$ Yukawa coupling in
\eetotth process can be measured to $6-8\%$ accuracy with integral
luminosity $1000~fb^{-1}$ at an $e^+ e^-$ linear collider (LC)
with $\sqrt{s}=1~TeV$\cite{tth1,Baer}. The accurate predictions
for the process \eetotth at linear colliders in $e^+e^-$ collision
mode have been intensively discussed in many
literatures\cite{tth2}-\cite{CSLi}. Chong-Xing Yue, {\cal et al.,}
studied the \eetotth process in the $LH$ and $LHT$ model at
$ILC$\cite{Yue,Yue1}. They found that in the parameter space
preferred by the electroweak precision data in the $LH$ model($f=1
\sim 2~TeV$, $c=0 \sim 0.5$, $c'=0.62 \sim 0.73$)\cite{Casaki},
the absolute value of the relative correction
$\delta\sigma/\sigma^{SM}$ can be larger than $5\%$, while in the
$LHT$ model as long as $f\leq1~TeV$ and $c_{\lambda}=0.1 \sim
0.9$, the value of $|\delta\sigma/\sigma^{SM}|$ can be larger than
$7\%$. That means in these parameter space the $LH/LHT$ model
effects might be observed in the future $ILC$ experiment. Except
the $e^+e^-$ collision mode, an $e^+e^-$ LC can also be operated
as a $\gamma\gamma$ collider. This is achieved by using Compton
backscattered photons in the scattering of intense laser photons
on the initial $e^+e^-$ beams. Generally $e^+e^-$ collider has the
advantage that the luminosity is higher than $\gamma\gamma$
collider, for example, ${\cal L}_{\gamma\gamma} \sim 0.15-0.2
~{\cal L}_{e^+e^-}$ or even $ 0.3-0.5 ~{\cal L}_{e^+e^-}$(through
reducing emittance in the damping rings)\cite{Telnov}, but the
polarization technique for photon is much simpler than electron
and the LC with continuous colliding energy spectrum of
$\gamma\gamma$ will be helpful to pursue new particles. Therefore,
LC can provide anther possibility to measure precisely the $t-\bar
t-h^0$ coupling in $\gamma\gamma$ collision mode. Similar with the
study on the process \eetotth at LC, the evaluation of radiative
corrections to the process \rrtth is also significant for the
accurate experimental measurements of $t-\bar t-h^0$ Yukawa
coupling at LC. In the Ref.\cite{ChenHui}, the calculations of the
cross sections for \rrtth and \eetth process including NLO QCD and
one-loop electroweak corrections in the $SM$ were presented.

\par
Due to the fact that it is speculated that the Yukawa coupling
between top-quarks and Higgs boson is theoretically sensitive to
the $LH$ and $LHT$ contributions, and the $t \bar{t} h^0$
associated productions may be favorable for probing these little
Higgs models. In this paper we study the reach of the ILC
operating in $\gamma\gamma$ collision mode to probe the $LH$ and
$LHT$ model in the process \eetth at the QCD next-to-leading
order. The paper is organized as follows. In Sec. 2 we give a
brief review of the $LH$ and $LHT$ model. In Section 3, we present
the notations and analytical calculation of the process \eetth
including the QCD NLO radiative corrections. The numerical result
and discussions are presented in Section 4. Finally the
conclusions are given.

\par
\section{Related theory of the $LH$ and $LHT$ models }
\par
Before our calculations, we will briefly recapitulate the $LH$ and
$LHT$ model which are relevant to the analysis in this work. For
the detailed description of these two models, one can refer to
Refs.\cite{LH5,LHT}. The littlest Higgs($LH$) model is based on
the $SU(5)/SO(5)$ non-linear sigma model\cite{littlest7}. In this
model, the SM fermions acquire their masses via the usual Yukawa
interactions. However, to cancel the large quadratic divergence in
the Higgs boson mass due to the heavy top quark Yukawa interaction
in the $SM$, a pair of new colored weak singlet Weyl fermions
$\tilde{t}$ and $\tilde{t}^{\prime c}$ is required in addition to
the usual third family weak doublet $q_3=(t_3,b_3)$ and weak
singlet $u^{\prime c}_3$, where $u^{\prime c}_3$ and
$\tilde{t}^{\prime c}$ are the corresponding right-handed
singlets. And the third family $SM$ quark doublet is replaced by a
chiral triplet field $\chi=(b_3,t_3,\tilde{t})$. The Lagrangian
generating the Yukawa couplings between pseudo-Goldstone bosons
and the heavy vector-like fermion pair in the $LH$ model is taken
the form as\cite{littlest7}:
\begin{eqnarray}
{\cal L}_Y &=& \frac{1}{2}\lambda_1 f
\epsilon_{ijk}\epsilon_{xy}\chi_i \Sigma_{jx}\Sigma_{ky}u^{\prime
c}_3+ \lambda_2f \tilde{t}\tilde{t}^{\prime c}+h.c.  \label{lag}
\end{eqnarray}

where $\epsilon_{ijk}$ and $\epsilon_{xy}$ are antisymmetric
tensors. $i$, $j$, $k$ run through 1, 2, 3 and $x$, $y$ run
through 4, 5. $\lambda_1$ and $\lambda_2$ are the coupling
constants. By expanding above Lagrangian, we get the physical
states of the top quark $t$ and a new heavy-vector-like quark $T$,
and obtain the usual mass result for the eigenvalues corresponding
to the top quark $t$ and the heavy top $T$ which are up to order
${\cal O} (v/f)$:
\begin{equation}
m_t  = \frac{\lambda_1 \lambda_2}{\sqrt{\lambda_1^2 +
\lambda_2^2}}\; v\;, \;\;\; m_T = \sqrt{\lambda_1^2 +
\lambda_2^2}\; f.\label{mt}
\end{equation}

\par
From the Lagrangian shown in Eq.(\ref{lag}), the couplings in the
$LH$ model concerned in the calculation of \rrtth process can be
expressed as:
\begin{eqnarray}\label{httLH}
g_{t\bar{t}h}^{LH} &=& - i \frac{m_t}{v} \left[ 1 -
\frac{1}{2}s_0^2+\frac{v}{f} \frac{s_0}{\sqrt{2}} - \frac{2
v^2}{3f^2} + \frac{v^2}{f^2}c_{\lambda}^{2}
( 1 + c_{\lambda}^{2})\right]\\
g_{T\bar{T}h}^{LH} &=& - i
\frac{\lambda_1^2}{\sqrt{\lambda_1^2+\lambda_2^2}} \left( 1 +
c_{\lambda}^{2} \right) \frac{v}{f},
\end{eqnarray}
where $v$ is one of the vacuum expectation
values($v=(\sqrt{2}G_{F})^{-1/2} = 246.22~GeV$), $c_{\lambda}$ is
define as $c_{\lambda}=\frac{\lambda_1}
{\sqrt{\lambda_1^2+\lambda_2^2}}$\cite{ref14,ref15}($\lambda_1$
and $\lambda_2$ are the Yukawa coupling parameters), $s_0$ is the
scalar mixing angle of Higgs fields, $s_0 \simeq 2
\sqrt{2}\frac{v'}{v} =\frac{xv}{\sqrt{2}f} \sim \mathcal{O}(v/f)$,
where we define $x\equiv4f\frac{v'}{v^2}$\cite{pheno5}.

\par
Recently, the symmetry structure of the the $LH$ model was
enlarged by introducing an additional discrete symmetry, T-parity,
in analogy to the R-parity in the minimal supersymmetric standard
model($MSSM$)\cite{LHT}. The T-parity interchanges the two
subgroups $[SU(2)\times U(1)]_1$ and $[SU(2)\times U(1)]_2$ of
$SU(5)$. Due to T-parity, the new gauge bosons do not mix with the
$SM$ gauge bosons and thus the new particles don't generate
corrections to precision electroweak observables at tree level.
The top quark sector contains a T-even and T-odd partner, with the
T-even one mixing with top quark and cancelling the quadratic
divergence contribution of top quark to Higgs boson mass. The mass
of the T-even partner (denoted as $T$) is the same as shown in
Eq.(\ref{mt}), while the mass of the T-odd partner (denoted as
$T_{-}$) is given by
\begin{eqnarray}
 m_{T_-}= \lambda_2 f,
\end{eqnarray}

\par
The mixing of $T$-quark with the top quark will alter the $SM$ top
quark couplings, and the relevant couplings in the $LHT$ model
using in our calculation are given as
\begin{eqnarray}\label{htt}
g_{t\bar{t}h}^{LHT}&=&-i\frac{m_{t}}{v}\left[1-\left(\frac{3}{4}-c_{\lambda}^{2}+
c_{\lambda}^{4}\right)\frac{v^{2}}{f^{2}}\right],\\ \label{htt1}
g_{T\bar{T}h}^{LHT}&=&{i \frac{m_t c_\lambda s_\lambda}{ f }},
\end{eqnarray}
where $s_{\lambda}=\sqrt{1-c_{\lambda}^{2}}$, and the Feynman
rules for the third generation quarks-gluon($\gamma$) couplings in
both $LH$ and $LHT$ models, have the same forms as the $gf \bar
f(\gamma f \bar f)$ couplings in the $SM$.

\par
Moreover, the top quark mass is already obtained by experiment,
then we can get the parameter relation from Eq. (\ref{mt}) as
deduced in Ref.\cite{littlest7}
\begin{eqnarray}
\label{l1l2}\frac{1}{\lambda_1^2}+\frac{1}{\lambda_2^2}\approx
\frac{v^2}{{m_t}^2} \approx 2.
\end{eqnarray}

\par
\section{Analytical calculations}
\par
\subsection{LO calculations of the \rrtth subprocess}
\par
\begin{figure}[hbtp]
\vspace*{-1cm} \centerline{ \epsfxsize = 12cm \epsfysize = 3.5cm
\epsfbox{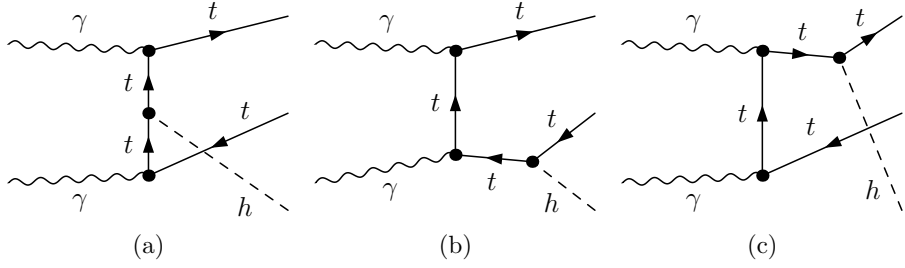}}  \vspace*{0cm}\caption{ The lowest order
diagrams for the $\gamma\gamma \to t\bar{t}h^0$ subprocess. }
\label{born}
\end{figure}

We denote the subprocess \rrtth as
\begin{equation}
\gamma(p_1,\lambda_1)+\gamma(p_2,\lambda_2) \to
t(k_1)+\bar{t}(k_2)+h^0(k_3).
\end{equation}

where the four-momenta of incoming electron and positron are
denoted as $p_1$ and $p_2$, and the four-momenta of outgoing
top-quark, anti-top-quark and Higgs boson are represented as
$k_1$, $k_2$ and $k_3$ respectively, $\lambda_{1,2}$ are the
polarizations of incoming photons. The tree-level t-channel
Feynman diagrams  are shown in Fig.\ref{born}, the u-channel with
the exchange of the two incoming photons are not shown. There
Higgs boson radiates from the internal or external top-quark
lines, so the cross section should be proportional to factor $g_{t
\bar t h}^2$. Consequently, this process can be used to probe the
$t-\bar t-h^0$ Yukawa coupling directly. \vskip 1cm

\par
The amplitudes of the corresponding t-channel Feynman diagrams
(shown in Fig.\ref{born}(a-c)) for the subprocess $\gamma\gamma
\to t\bar{t}h^0$ are expressed as
\begin{eqnarray}
{\cal M}^{(a)}_t=-\frac{e^3
Q_t^2m_t}{2m_W\sin\theta_W}\frac{1}{(k_1-p_1)^2-m_t^2}\frac{1}{(p_2-k_2)^2-m_t^2} \cdot ~~~~\nb \\
\times\bar{u}(k_1)\rlap/{\epsilon}(p_1,\lambda_1)(\rlap/{k}_1-\rlap/{p}_1+m_t)(\rlap/{p}_2-\rlap/{k}_2+m_t)
\rlap/{\epsilon}(p_2,\lambda_2)v(k_2),
\end{eqnarray}
\begin{eqnarray}
{\cal M}^{(b)}_t=-\frac{e^3
Q_t^2m_t}{2m_W\sin\theta_W}\frac{1}{(k_1-p_1)^2-m_t^2}\frac{1}{(k_2+k_3)^2-m_t^2} \cdot ~~~~\nb \\
\times\bar{u}(k_1)\rlap/{\epsilon}(p_1,\lambda_1)(\rlap/{k}_1-\rlap/{p}_1+m_t)\rlap/{\epsilon}(p_2,\lambda_2)
(-\rlap/{k}_2-\rlap/{k}_3+m_t)v(k_2),
\end{eqnarray}
\begin{eqnarray}
{\cal M}^{(c)}_t=-\frac{e^3
Q_t^2m_t}{2m_W\sin\theta_W}\frac{1}{(k_1+k_3)^2-m_t^2}\frac{1}{(p_2-q_2)^2-m_t^2} \cdot ~~~~\nb \\
\times\bar{u}(k_1)(\rlap/{k}_1+\rlap/{k}_3+m_t)\rlap/{\epsilon}(p_1,\lambda_1)(\rlap/{p}_2-\rlap/{k}_2+m_t)
\rlap/{\epsilon}(p_2,\lambda_2)v(k_2),
\end{eqnarray}
where $Q_t=2/3$ and the corresponding amplitudes of the u-channel
Feynman diagrams of the subprocess $\gamma\gamma \to t\bar{t}h^0$
can be obtained by exchanging
$\gamma(p_1,\lambda_1)\leftrightarrow \gamma(p_2,\lambda_2)$.
\begin{eqnarray}
{\cal M}^{(a)}_u={\cal M}^{(a)}_t(p_1,\lambda_1 \leftrightarrow
p_2,\lambda_2),&&{\cal M}^{(b)}_u={\cal M}^{(b)}_t(p_1,\lambda_1
\leftrightarrow p_2,\lambda_2), \nb \\&& {\cal M}^{(c)}_u={\cal
M}^{(c)}_t(p_1,\lambda_1 \leftrightarrow p_2,\lambda_2).
\end{eqnarray}

\par
The total amplitude at the lowest order is the summation of the
above amplitudes.
\begin{equation}
{\cal M}_0=\sum_{i=a,b}^{c} \sum_{j=u}^{t}{\cal M}^{(i)}_{j}.
\end{equation}

The cross section of the subprocess \rrtth in unpolarized photon
collision mode at the tree-level can be obtained by integrating
over the phase space,
\begin{eqnarray}
\hat{\sigma}_0(\hat{s}) =  \frac{(2 \pi )^4
N_c}{4|\vec{p}_1|\sqrt{s}}\int {\rm d} \Phi_3 \overline{\sum_{{\rm
spin}}} |{\cal M}_{0}|^2,
\end{eqnarray}
where $N_c=3$ and $\vec{p}_1$ is the c.m.s. momentum of one
initial photon, $d\Phi_3$ is the three-body phase space element,
and the bar over summation recalls averaging over initial
spins\cite{data}.

\vskip 15mm
\subsection{Calculations of the QCD NLO corrections of the \rrtth subprocess}
\par
\begin{figure}[hbtp]
\vspace*{-1cm} \centerline{ \epsfxsize = 12cm \epsfysize = 8.5cm
\epsfbox{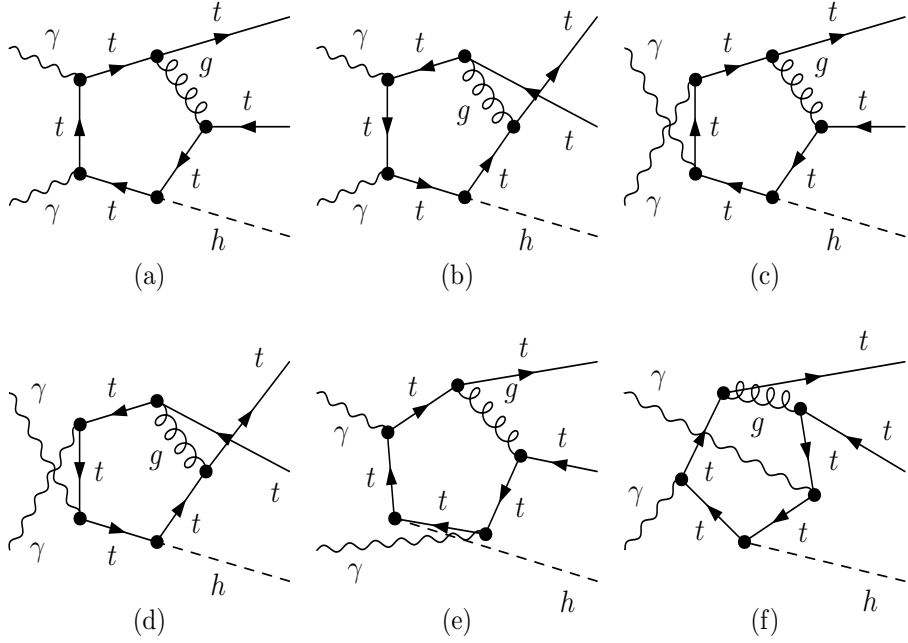}}  \vspace*{0cm}\caption{ The representative QCD
one-loop diagrams which produce amplitudes include five-point
tensor integrals of rank 4.} \label{box}
\end{figure}

The ${\cal O}(\alpha_{s})$ QCD NLO Feynman diagrams of the
subprocess \rrtth are generated by ${\it FeynArts}~3$ \cite{FA3}.
The QCD NLO Feynman diagrams can be divided into self-energy,
vertex, box, pentagon and counter term diagrams. We find there
exist the QCD one-loop diagrams which include
$T$-quark/$T_{\pm}$-quark in loops for the $LH$/$LHT$ model, but
the total contributions from these diagrams are vanished in both
models separately. The representative pentagon Feynman diagrams
which generate amplitudes including five-point integrals of rank 4
are shown in Fig.\ref{box}. The amplitude of the subprocess \rrtth
including virtual QCD corrections to ${\cal O}(\alpha_s)$ order
can be expressed as
\begin{equation}
{\cal M}_{QCD}={\cal M}_0+{\cal M}_{QCD}^{vir},
\end{equation}
where ${\cal M}_{QCD}^{vir}$ is the renormalized amplitude
contributed by the QCD one-loop Feynman diagrams, the QCD
renormalizations of top-quark wave function, mass and
$t-\bar{t}-h^0$ Yukawa coupling. There we define the relevant QCD
renormalization constants as
\begin{eqnarray}\label{RenConstant}
m_{t,0}=m_t+\delta m_{t(g)},~~ t_0^{L}=\left(1+\frac{1}{2}\delta
Z_{t(g)}^L\right)t^L, \nb   \\
t_0^{R}=\left(1+\frac{1}{2}\delta Z_{t(g)}^R\right)t^R,
~~g_{t\bar{t}h}^0=g_{t\bar{t}h}\left(1+\frac{\delta
m_{t(g)}}{m_t}\right).
\end{eqnarray}

In analogy to the calculation of the QCD renormalization constants
in Ref.\cite{ChenHui}, we adopt the on-mass-shell renormalization
condition to get the QCD contributed parts of the renormalization
constants, $\delta m_{t(g)}$ and $\delta Z_{t(g)}^{L,R}$.

\vskip 1cm

\par
The virtual QCD corrections contain both ultraviolet (UV) and soft
infrared (IR) divergences. We adopt the dimensional
regularization($DR$) to regularize the UV divergences in loop
integrals, and to isolate IR singularities. After renormalization
procedure, the virtual correction part of the cross section is
UV-finite. The IR divergences from the one-loop diagrams involving
virtual gluon can be cancelled by adding the soft real gluon
emission corrections by using the phase space slicing method
(PSS)\cite{PSS}. The real gluon emission process is denoted as
\begin{eqnarray}
\label{real gluon emission}
 \gamma(p_1,\lambda_1)+\gamma(p_2,\lambda_2) \rightarrow t(k_1)+\bar{t}(k_2)+h^0(k_3)+g(k),
\end{eqnarray}

where a real gluon radiates from the internal or external
top(anti-top) quark line. The phase space for \rrtthg process is
divided into two parts which behave soft and hard gluon emission
natures, respectively.
\begin{equation}
\Delta \hat{\sigma}_{real}^{QCD}=\Delta \hat{\sigma}_{
soft}^{QCD}+\Delta \hat{\sigma}_{hard}^{QCD}.
\end{equation}

\par
Finally the UV and IR finite total cross section of the subprocess
\rrtth including the ${\cal O}(\alpha_{s})$ QCD corrections is
obtained as
\begin{equation}\label{cs}
\hat{\sigma}^{QCD}  = \hat{\sigma}_0 + \Delta \hat{\sigma}^{QCD} =
\hat{\sigma}_0 +  \Delta \hat{\sigma}_{{\rm vir}}^{QCD} +
 \Delta \hat{\sigma}_{{\rm real}}^{QCD} = \hat{\sigma}_0(1 +
\hat{\delta}^{QCD}),
\end{equation}
where $\hat{\delta}^{QCD}$ is the QCD relative correction of order
${\cal O}(\alpha_s)$.

\vskip 10mm
\par
\subsection{Calculations of process \eetth }
\par
The hard photon beam of the $\gamma\gamma$ collider can be
obtained by using the laser back-scattering technique at $e^+e^-$
linear collider \cite{Com1,Com2,Com3}. For simplicity, in our
calculations we ignore the possible polarization for the incoming
electron and photon beams. We denote $\hat{s}$ and $s$ as the
center-of-mass energies of the $\gamma\gamma$ and $e^{+}e^{-}$
systems, respectively. After calculating the cross section
$\hat{\sigma}^{QCD}(\hat{s})$ for the subprocess \rrtth in
unpolarized photon collision mode, the total cross section at an
$e^{+}e^{-}$ linear collider can be obtained by folding
$\hat{\sigma}^{QCD}(\hat{s})$ with the photon distribution
function that is given in Ref.\cite{function},
\begin{equation}
\sigma_{tot}(e^+e^- \to \gamma\gamma \to t\bar th^0
,~s)=\int^{x_{max}}_{(2m_t+m_h)/\sqrt{s}} dz\frac{d{\cal
L}_{\gamma\gamma}}{dz} \hat{\sigma}^{QCD}(\hat{s}=z^2 s ).
 \end{equation}

The distribution function of photon luminosity $\frac{d{\cal
L}_{\gamma\gamma}}{d z}$ is expressed as
\begin{eqnarray}
\frac{d{\cal L}_{\gamma\gamma}}{dz}=2z\int_{z^2/x_{max}}^{x_{max}}
 \frac{dx}{x} f_{\gamma/e}(x)f_{\gamma/e}(z^2/x)
\end{eqnarray}

The energy spectrum of the back scattered photon in unpolarized
incoming $e^-\gamma$ scattering is given by
\begin{eqnarray}
\label{structure}
f_{\gamma/e}(x)=\frac{1}{D(\xi)}\left[1-x+\frac{1}{1-x}-
\frac{4x}{\xi(1-x)}+\frac{4x^{2}}{\xi^{2}(1-x)^2}\right],~~~(x<x_{max})
\end{eqnarray}

where the fraction of the energy of the incident electron carried
by the back-scattered photon $x$, is expressed as $x=2
\omega/\sqrt{s}$, and $x_{max}=2
\omega_{max}/\sqrt{s}=\xi/(1+\xi)$. For $x>x_{max}$,
$f_{\gamma/e}=0$. The function $D(\xi)$ is defined as
\begin{equation}
D(\xi)=(1-\frac{4}{\xi}-\frac{8}{\xi^{2}})\ln(1+\xi)+\frac{1}{2}+\frac{8}{\xi}-\frac{1}{2(1+\xi)^{2}}.
\end{equation}
We denote $m_{e}$ and $\omega_0$ as electron mass and laser-photon
energy respectively. The incoming electron energy is $\sqrt{s}/2$
and $\xi=\frac{2 \omega_0\sqrt{s}}{{m_e}^2}$. In our evaluation,
we choose $\omega_0$ such that it maximizes the backscattered
photon energy without spoiling the luminosity through $e^{+}e^{-}$
pair creation. Then we have ${\xi}=2(1+\sqrt{2})$, $x_{max}\simeq
0.83$, and $D(\xi) \approx 1.84$\cite{photon para}.

\par
\section{Numerical results and discussions}
\par
In this section, we present some numerical results for both the
\rrtth subprocess and \eetth parent process in the littlest Higgs
model and its extension model with T-parity(the $LH$ and $LHT$
model). In the numerical calculation, we take the input parameters
as follows\cite{data}
\begin{eqnarray}
\alpha_{{\rm ew}}(0)^{-1}=137.03599911,~~~
m_W = 80.403~GeV,~~~m_Z = 91.1876~GeV, \nb \\
m_t = 174.2~GeV, ~~~\alpha_s(m_Z) = 0.117620.
\end{eqnarray}

The mixing parameter $s_{0}$, which appears in the $LH$
coupling(see Eq.(\ref{httLH})), is of the order ${\cal O}(v/f)$.
We fix $s_{0}= \frac{v}{2 f}$(It is equivalent to
$x=\frac{1}{\sqrt{2}}$) in numerical evaluation, if there is no
other statement. Then we still have additional four free
$LH$/$LHT$ parameters ($f$, $c_{\lambda}$, $\sqrt{s}$, $m_h$)
involved in our numerical calculations. C. Csaki, etal., performed
a global fit to the precision data, and they found for generic
regions of the parameter space of little Higgs models the bound on
scale $f$ is several TeV, but there exist regions of parameter
space in which $f$ can be relaxed to $1-2~TeV$ depending on the
model variation and degree of tuning of model
parameters\cite{Casaki}. Considering the fact as shown in our
numerical results for the process \eetth, the corrections from the
$LH$ model are always less than $5\%$ when $f>2~TeV$. That means
only the symmetry breaking scales $f$ up to $2~TeV$ are accessible
in measuring the $LH$ model effects in $t\bar th^0$ associated
production. Then in the following numerical calculation in the
$LH$ model we constraint the value of the scale $f$ being in the
range of $1-2~TeV$. When it comes to the $LHT$ model, as the $SM$
gauge bosons can not mix with the new gauge bosons, and the
electroweak precision observables are not modified at tree level,
the symmetry breaking scale $f$ can be decreased to $500~GeV$,
which will lead to rich phenomenology of the $LHT$ model in
present and future high energy experiments. In this work we take
the QCD renormalization scale $\mu$ being $(2 m_t+m_h)/2$,
$c_{\lambda}\in[0.1,0.9]$ and the running strong coupling
$\alpha_s(\mu^2)$ being at the two-loop level($\overline{MS}$
scheme) with five active flavors.

\par
\begin{figure}[htb]
\centering
\includegraphics[scale=0.38,bb=22 26 512 413]{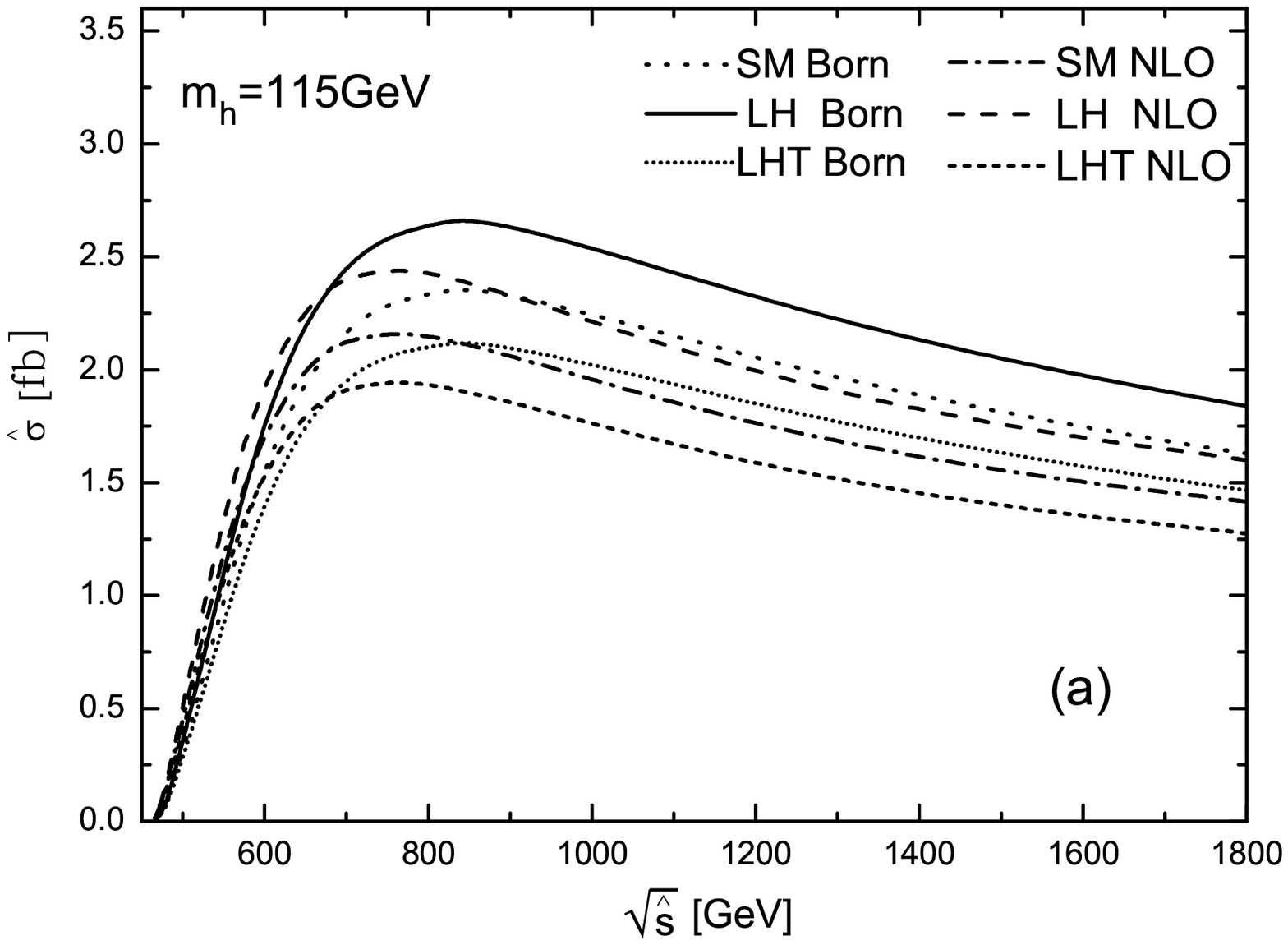}
\includegraphics[scale=0.38,bb=22 26 512 413]{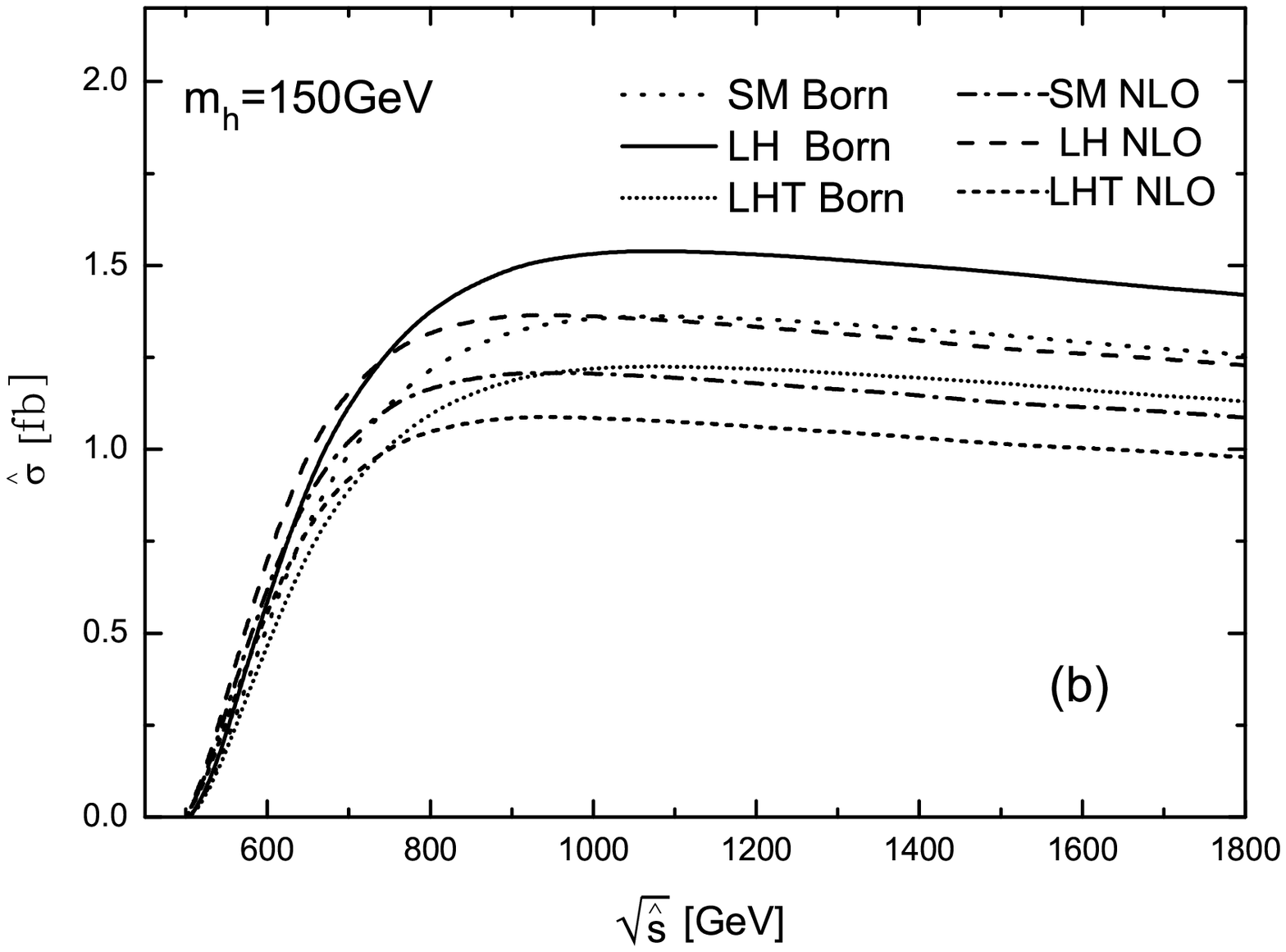}
\includegraphics[scale=0.38,bb=22 26 512 413]{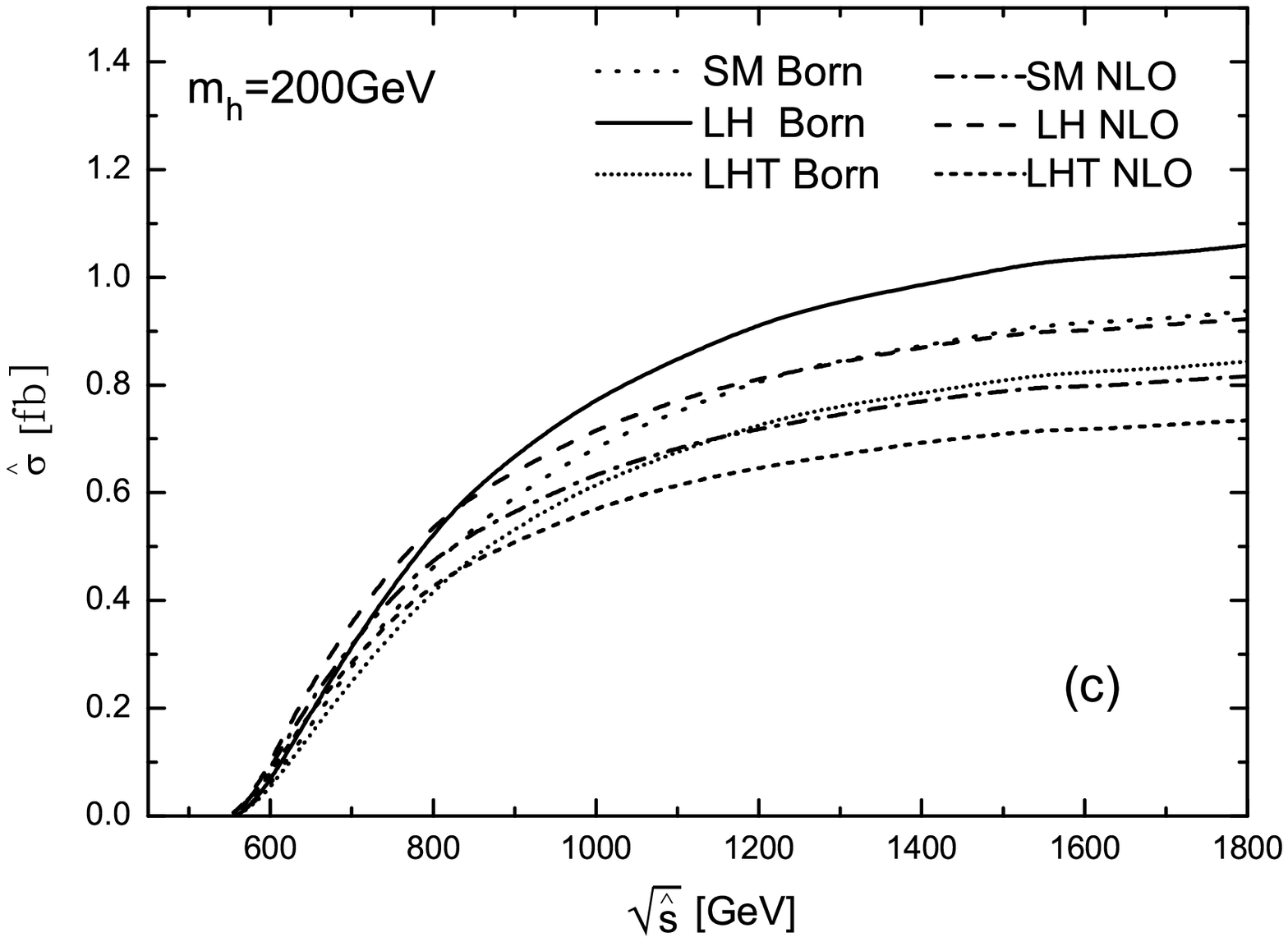}
\caption{ The Born and the QCD NLO corrected cross sections of the
process \rrtth as the functions of the c.m.s. energy
$\sqrt{\hat{s}}$ of $\gamma\gamma$ collision with $m_h=115$, $150$
and $200~GeV$ respectively, in the conditions of $f = 1~TeV$ and
$c_{\lambda}^2=0.8$. (a) for $m_h=115~GeV$, (b) for $m_h=150~GeV$,
(c) for $m_h=200~GeV$.} \label{Fig3}
\end{figure}

The numerical results for the cross sections of \rrtth including
QCD NLO radiative corrections versus $\gamma\gamma$ colliding
energy $\sqrt{\hat{s}}$, are plotted in Figs.\ref{Fig3}(a-c) with
$m_h=115$, $150$ and $200~GeV$ separately, where we take $f=
1~TeV$ and $c_{\lambda}^2 = 0.8$. The curves correspond to the
tree-level and QCD NLO corrected cross sections in the frameworks
of the $SM$, $LH$ and $LHT$ model respectively, with
$\sqrt{\hat{s}}$ running from the value little larger than the
threshold $2 m_t+m_h$ to $1.8~TeV$. Figs.\ref{Fig3}(a-c) show that
the QCD corrections can increase (when $\sqrt{\hat{s}} < 650~GeV$)
or decrease(when $\sqrt{\hat{s}}
> 900~GeV$) the tree-level cross sections of subprocess \rrtth. As indicated
in Fig.\ref{Fig3}(a), the curves for $m_h=115~GeV$ increase
rapidly to their maximal cross section values, when the $\gamma
\gamma$ colliding energy $\sqrt{\hat{s}}$ varies from threshold to
the corresponding position of peak. As depicted in
Fig.\ref{Fig3}(b) with $m_h=150~GeV$, all curves have platforms
when $\sqrt{\hat{s}}$ is larger than $1~TeV$. In Fig.\ref{Fig3}(c)
both Born and one-loop QCD corrected cross sections increase
slowly in our plotted range of $\sqrt{\hat{s}}$. From
Figs.\ref{Fig3}(a-c), we can also find that the tree-level and the
NLO QCD corrected cross sections in the $LH$ model, is always
larger than those in the other two models, while the cross section
of the $LHT$ model is the smallest one among all of the three
models.

\par
\begin{figure}[htb]
\centering
\includegraphics[scale=0.4,bb=22 26 512 413]{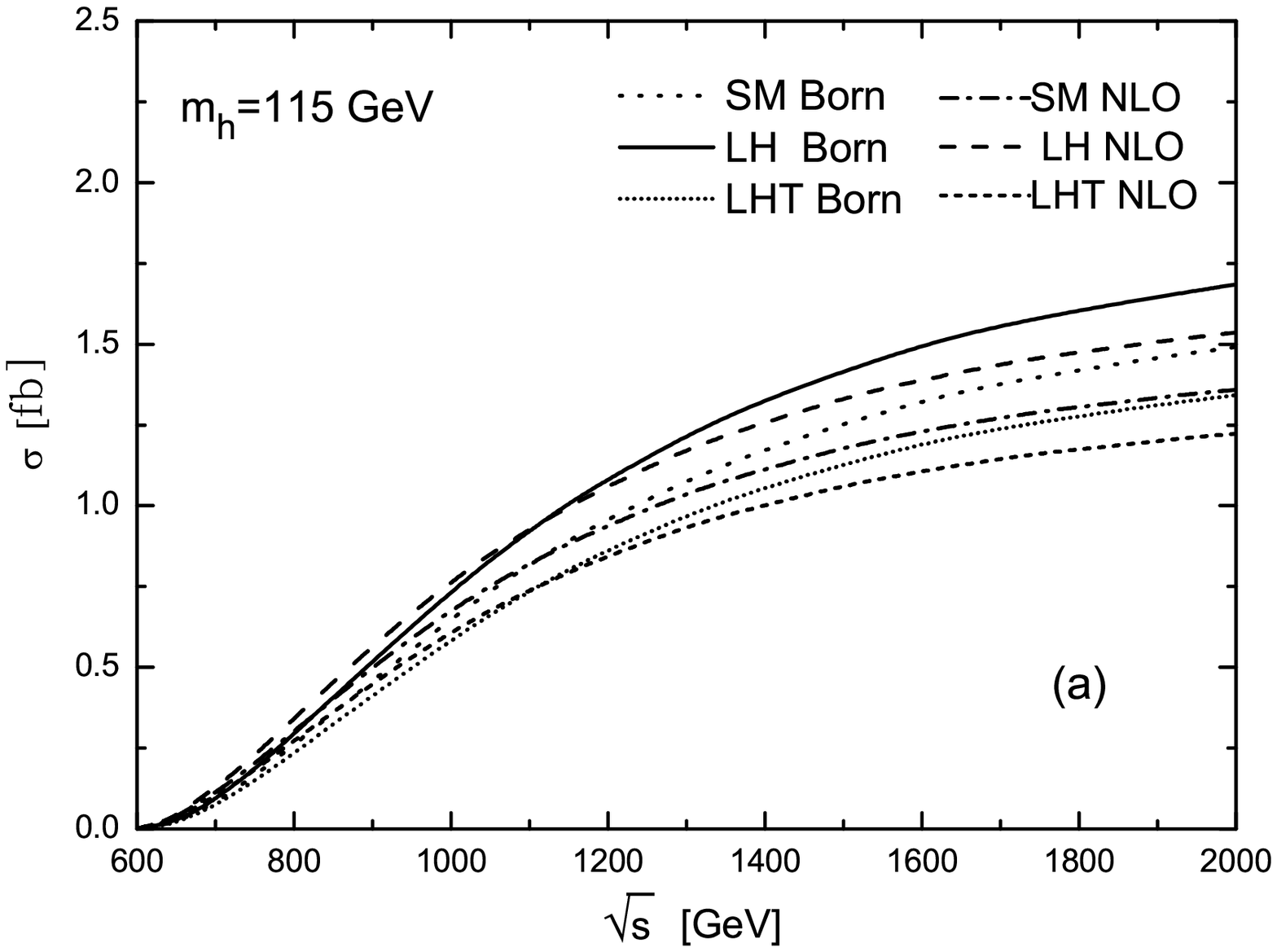}
\includegraphics[scale=0.4,bb=22 26 512 413]{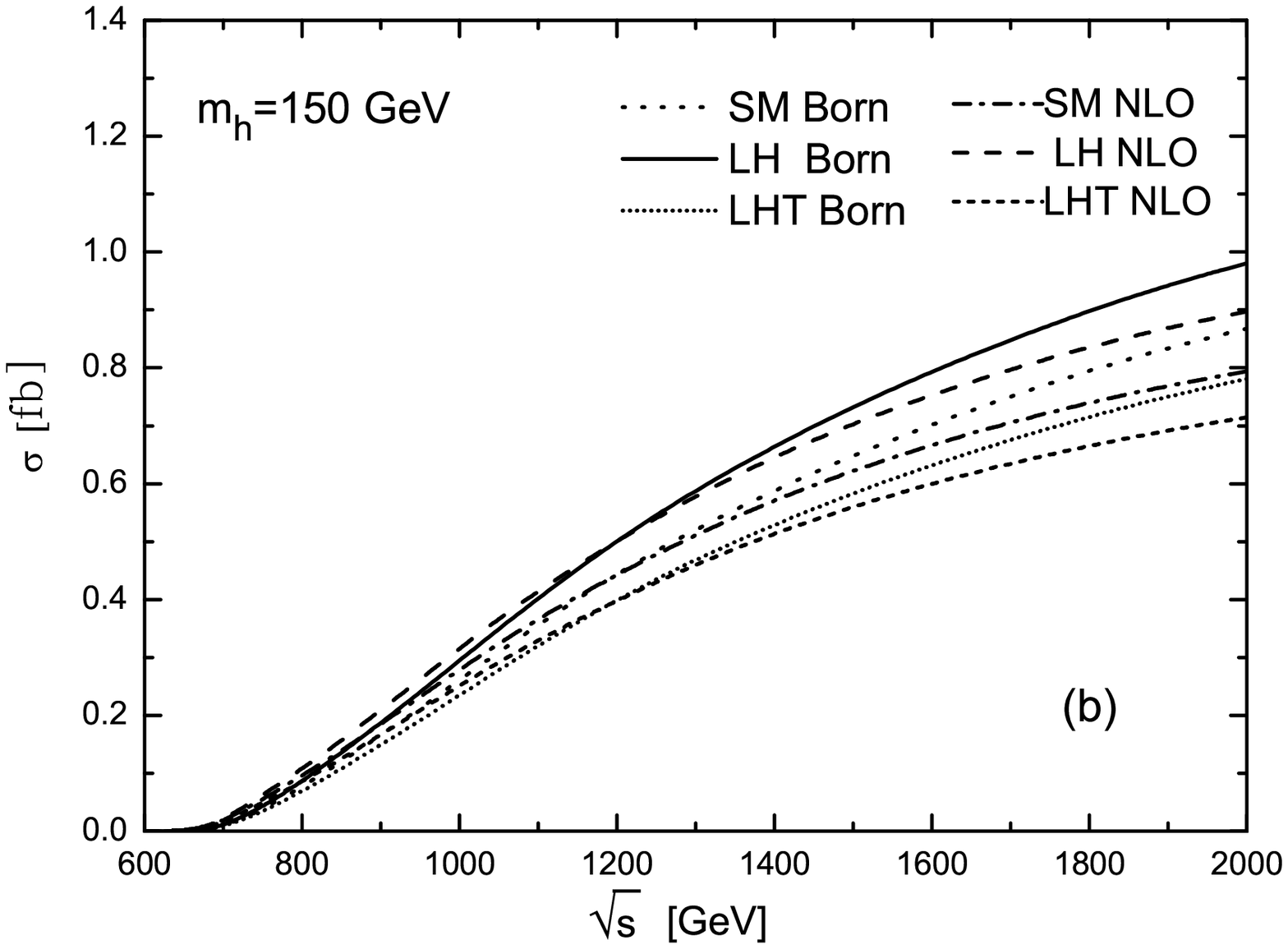}
\includegraphics[scale=0.4,bb=22 26 512 413]{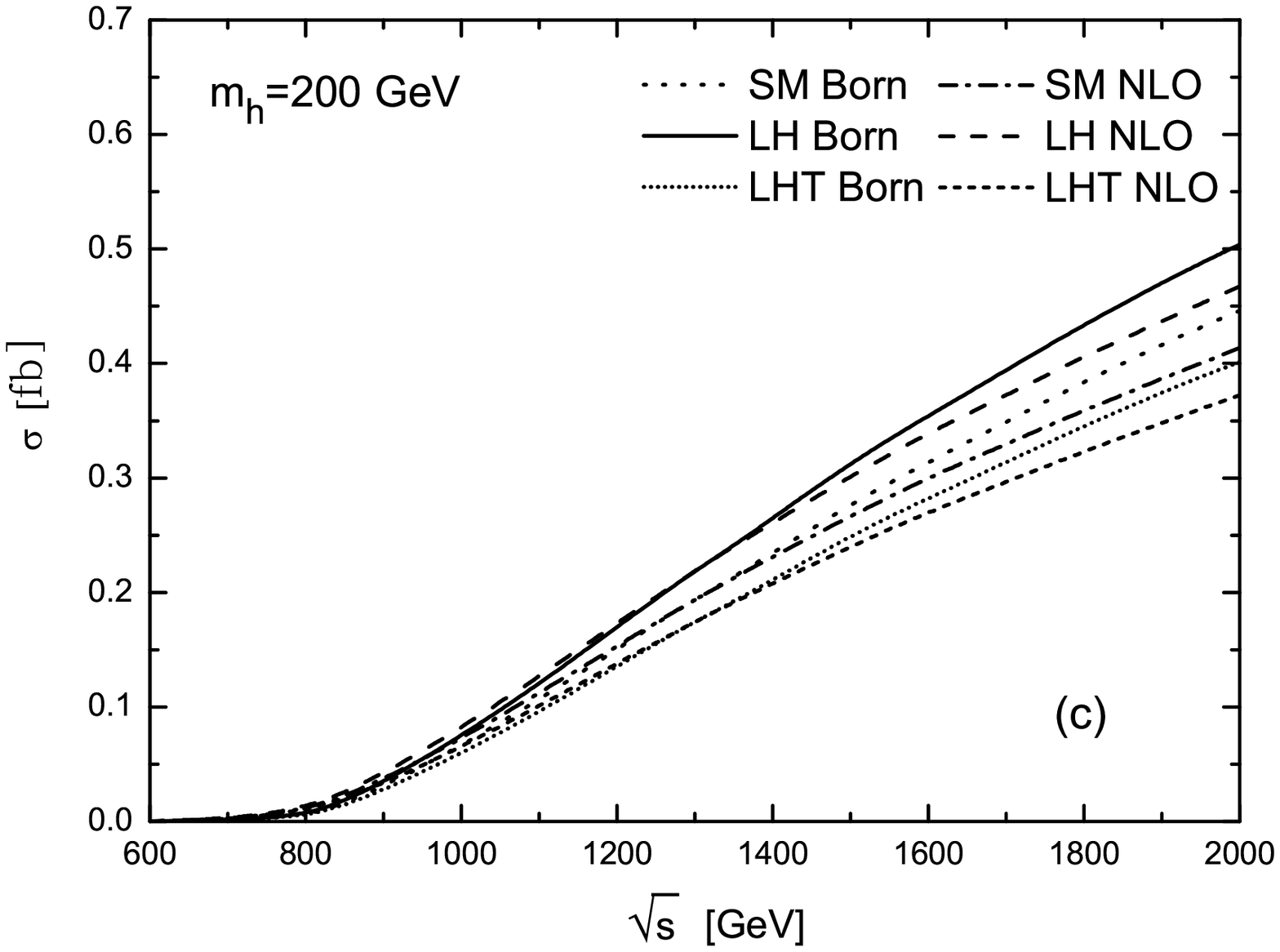}
\caption{ The Born and QCD NLO corrected cross sections for the
parent process \eetth as the functions of c.m.s. energy $\sqrt{s}$
with $m_h=115$, $150$, $200~GeV$ respectively, where $f=1~TeV$ and
$c_{\lambda}^{2}=0.8$. (a) is for $m_h=115~GeV$, (b) for
$m_h=150~GeV$, (c) for $m_h=200~GeV$. } \label{Fig4}
\end{figure}

The tree-level and the QCD NLO corrected cross sections for the
parent process \eetth in the $SM$, $LH$ and $LHT$ model as the
functions of the $e^+e^-$ colliding energy $\sqrt{s}$ in the
conditions of $f= 1~TeV$ and $c_{\lambda}^2 = 0.8$, are plotted in
Figs.\ref{Fig4}(a-c) for $m_h=115~GeV$, $150~GeV$ and $200~GeV$
separately. As shown in the figures, both Born and QCD NLO
corrected cross sections for each model go up with the increment
of $\sqrt{s}$, and the QCD NLO radiative corrections for different
value choices of $m_h$ can reduce or increase the Born cross
sections in the plotted $\sqrt{s}$ range. The tendencies of all
the curves in Figs.\ref{Fig4}(a), (b) and (c) are similar. The
cross section including QCD NLO corrections in the $LH$($LHT$)
model with $m_h=115~GeV$, can reach $1.5~fb(1.2~fb)$, and if we
assume the integral luminosity of an $e^+e^-$ linear collider
${\cal L}_{e^+e^-}=1000~fb^{-1}$, we can accumulate about
$1.5(1.2) \times 10^3$ $t \bar t h^0$ production events, thus it
will be helpful in hunting for the $LH$/$LHT$ signals and the
study of the Yukawa coupling.

\par
\begin{figure}[htb]
\centering
\includegraphics[scale=0.4,bb=22 26 512 413]{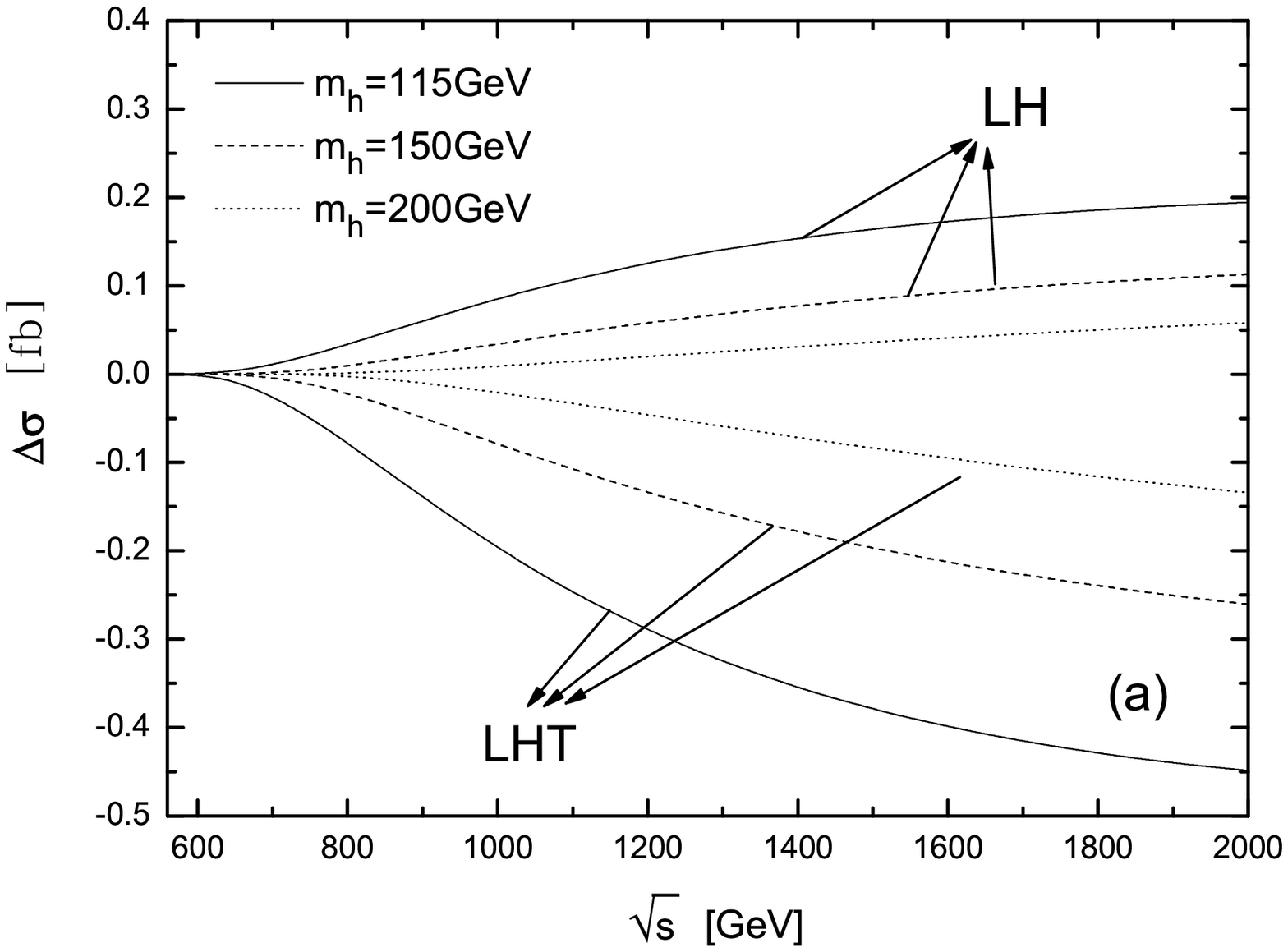}
\includegraphics[scale=0.4,bb=22 26 512 413]{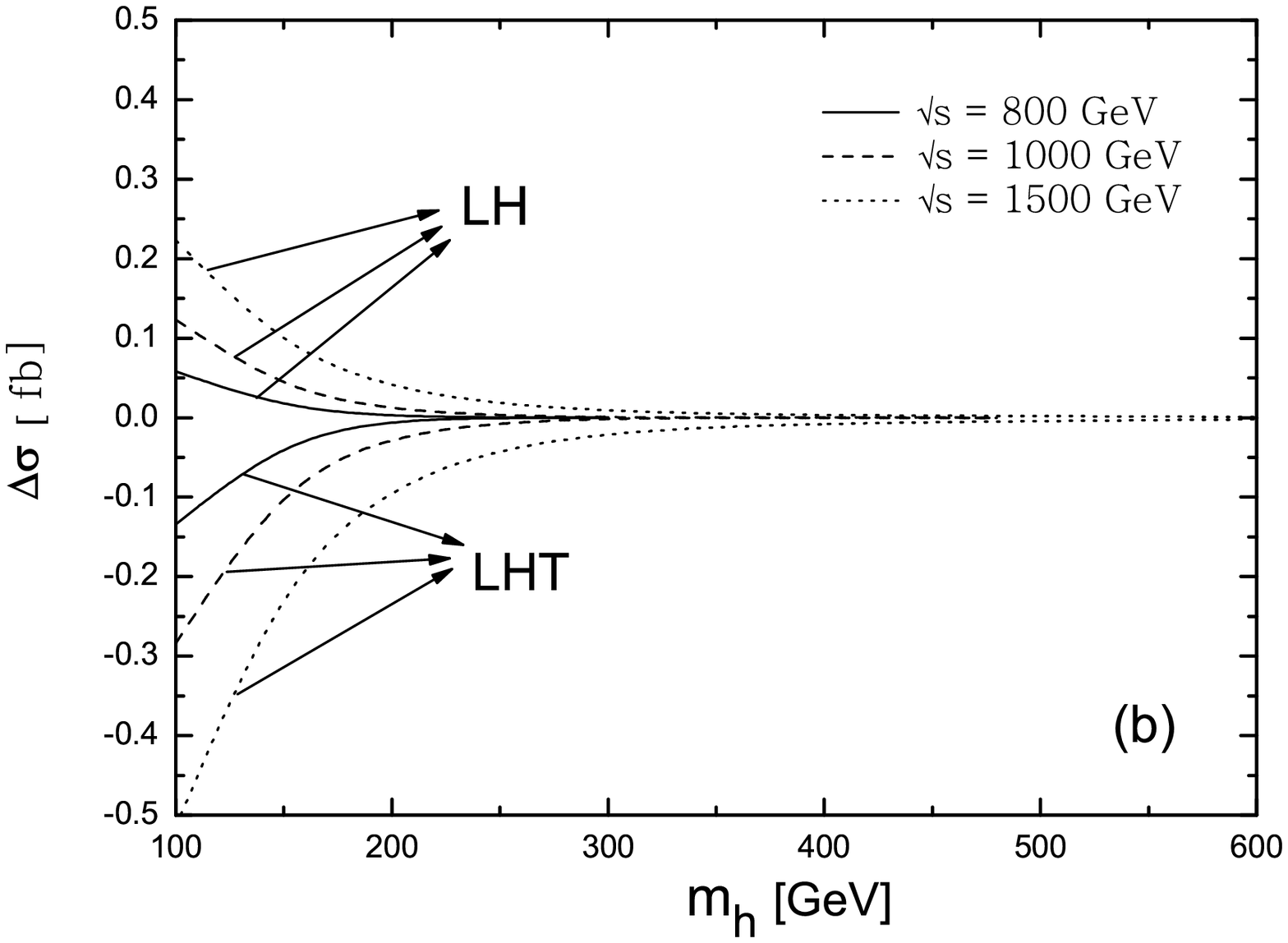}
\caption{ (a) The $\Delta\sigma_{LH,LHT} $ dependence on
$\sqrt{s}$ with $c_{\lambda}^{2}= 0.8$ and $f = 1(0.5)~TeV$(in the
$LH(LHT)$ model) for the process \eetth via $\gamma\gamma$
collision mode. (b) The $\Delta\sigma_{LH,LHT} $ dependence on
$m_h$ with $c_{\lambda}^{2}= 0.8$ and $f = 1(0.5)~TeV$(for the
$LH(LHT)$ model) for the process \eetth via $\gamma\gamma$
collision mode. } \label{Fig5}
\end{figure}

To illustrate the deviations of the cross sections in the
$LH$/$LHT$ model for the process \eetth from the $SM$ predictions,
we plot $\Delta\sigma_{LH,LHT}(
\equiv\sigma_{NLO}^{LH,LHT}-\sigma_{NLO}^{SM})$ as the functions
of $e^+e^-$ colliding energy $\sqrt{s}$ with the conditions of
$c_{\lambda}^{2}= 0.8$ and $f = 1(0.5)~TeV$(in the $LH$($LHT$)
model) in Fig.\ref{Fig5}(a). The solid, dashed and dotted lines
are for $m_h=115~GeV$, $150~GeV$ and $200~GeV$, respectively. For
each line type, the upper line is for the case in the $LH$ model,
while the lower one is in the $LHT$ model. From the figure, we can
see that the absolute value of the cross section deviation
$\Delta\sigma_{LH,LHT}$ raises with either the decrement of Higgs
mass or the increment of $e^+e^-$ c.m.s energy $\sqrt{s}$. The
deviations of the cross sections in the $LH$/$LHT$ model for the
process \eetth as the functions of the Higgs boson mass $m_h$ in
the same conditions as in Fig.\ref{Fig5}(a) are shown in
Fig.\ref{Fig5}(b). We can see from Fig.\ref{Fig5}(b) that for the
curves of $\sqrt{s}=1000~GeV$, $1500~GeV$, the absolute deviations
of the cross sections $|\Delta\sigma|$ can be larger than
$0.05~fb$ when $m_h\in [100~GeV,~150~GeV]$(in the $LH$ model) and
$m_h\in[100~GeV,~175~GeV]$(in the $LH$ model), and the effects
could be observable in experiment. But when $m_h$ is larger than
$300~GeV$, the effects from the $LH/LHT$ model become to be very
small and are not sensitive to the Higgs boson mass.
\vspace{0.5cm}

\par
\begin{figure}[htb]
\centering
\includegraphics[scale=0.35]{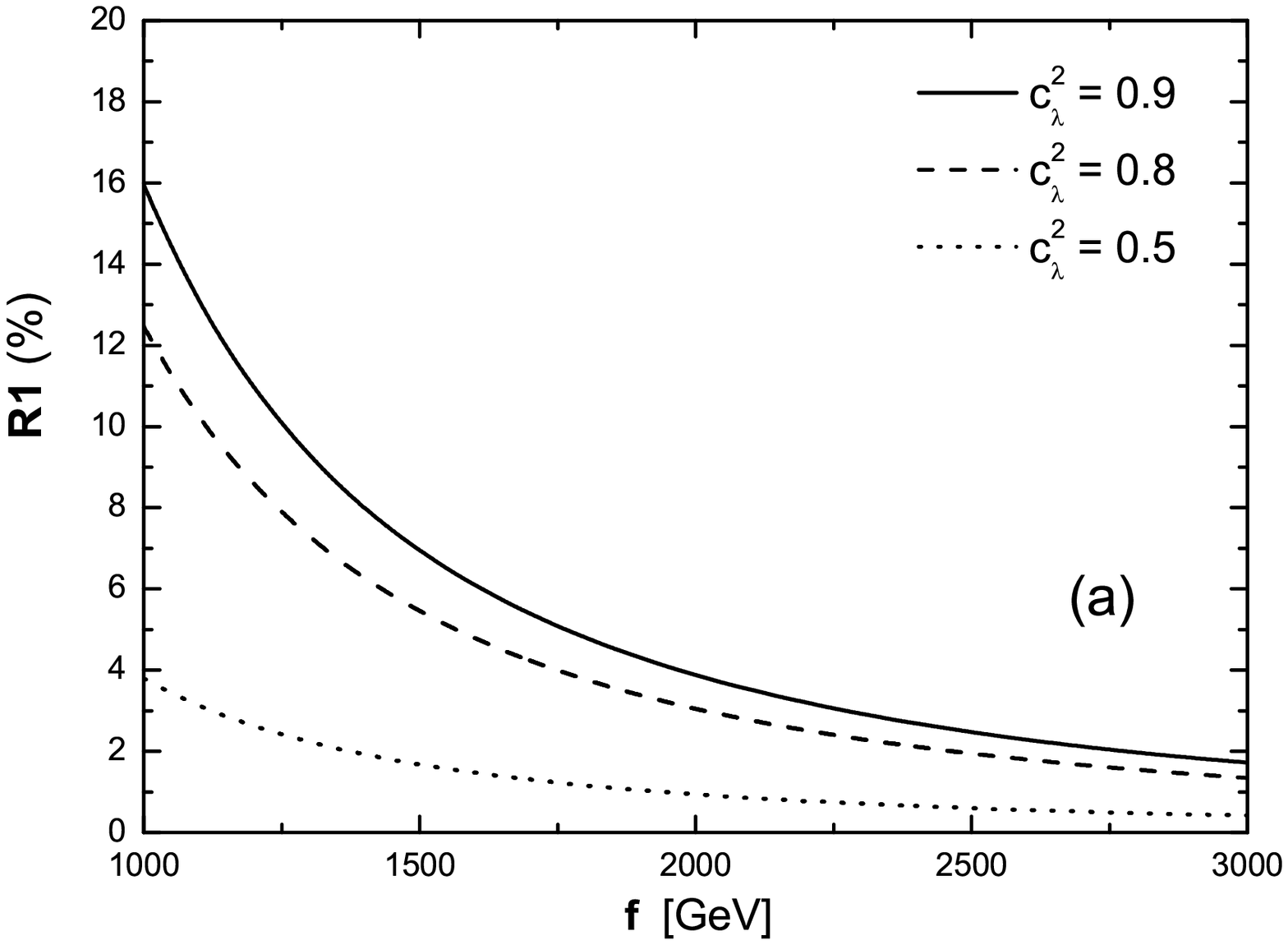}
\includegraphics[scale=0.35]{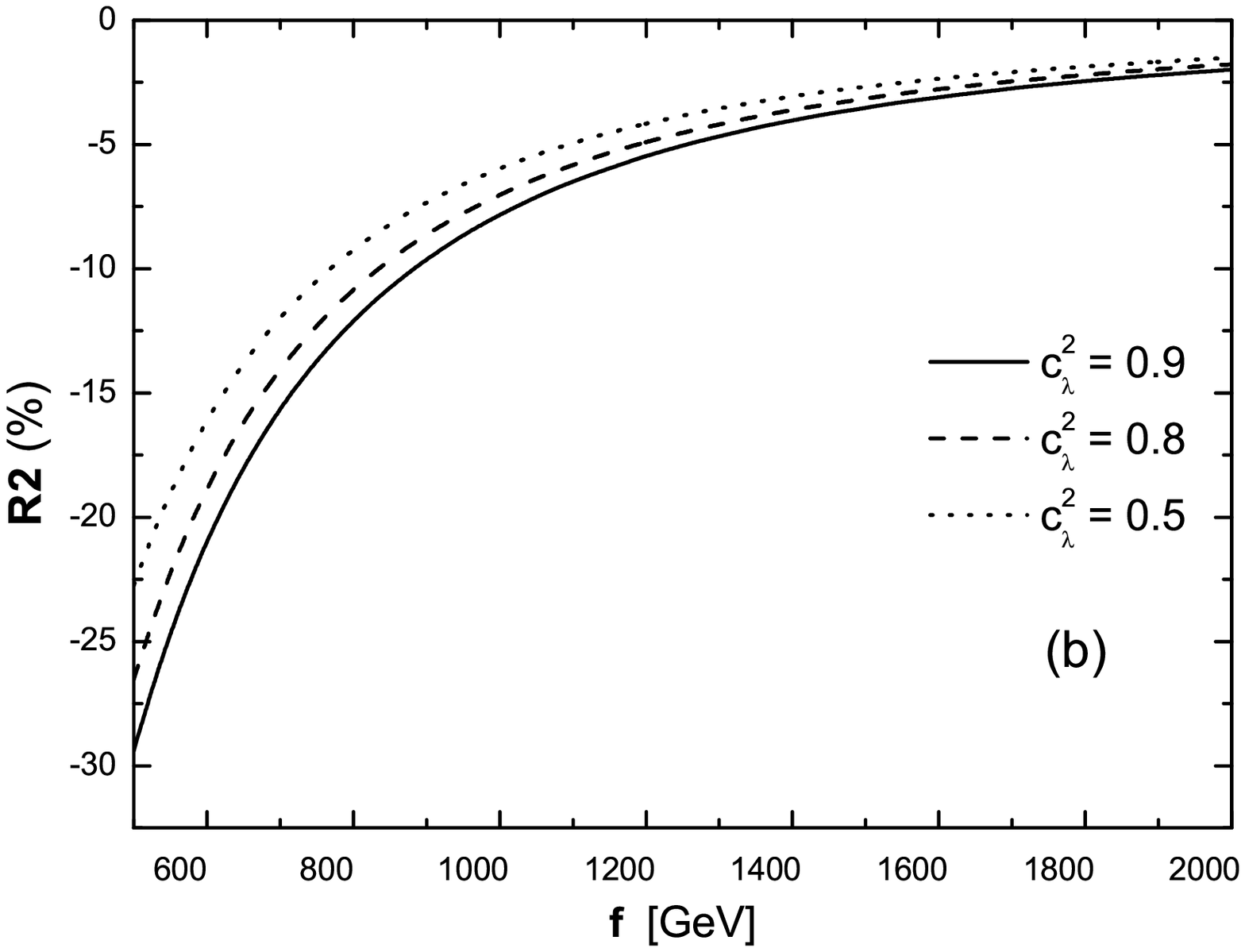}
\caption{ The relative parameters $R_1$ and $R_2$ for the process
\eetth as the functions of the symmetry breaking scale $f$ for
three different value choices of the mixing parameter
$c_{\lambda}$, with $m_h=115~GeV$ and $\sqrt{s}=800~GeV$. (a) for
$LH$ model and (b) for $LHT$ model} \label{Fig6}
\end{figure}

In general, the extra contribution of the $LH$ or $LHT$ model to
the cross section of the process \rrtth is proportional to a
factor of $1/f^{2}$. In order to describe the $LH$/$LHT$ effects
on the production cross section, we define the relative deviation
parameters
$R_{1}=\frac{\sigma_{NLO}^{LH}-\sigma_{NLO}^{SM}}{\sigma_{NLO}^{SM}}$
for the $LH$ model and
$R_{2}=\frac{\sigma_{NLO}^{LHT}-\sigma_{NLO}^{SM}}{\sigma_{NLO}^{SM}}$
for the $LHT$ model, and depict $R_1$ and $R_2$ as the functions
of symmetry breaking scale $f$ in Figs.\ref{Fig6}(a) and (b)
separately. In Figs.\ref{Fig6}(a,b) we take $m_h=115~GeV$,
$\sqrt{s}=800~GeV$, the mixing parameter $c_{\lambda}^2=0.5$,
$0.8$, $0.9$, and $f$ being in the range of $[1~TeV,~3~TeV]$ for
the $LH$ model, and $f \in [500~GeV,~2~TeV]$ for the $LHT$ model
respectively. From Fig.\ref{Fig6}(a), we can see that the relative
deviation parameter $R_1$ falls as $f$ increases, and when
$c_{\lambda}^{2} \geq 0.8$, the values of $R_{1}$ are larger than
$5\%$ in the range of symmetry breaking scale $f<1.5~TeV$, which
might be detected in the future $LC$ experiments. Since the
experimental constraint on symmetry breaking scale $f$ of the
$LHT$ model can be lower than $1~TeV$, the absolute value of
relative deviation parameter $R_{2}$ is generally larger than that
in the $LH$ model with the $f$ in the range of $[500~GeV,~1~TeV]$.
Similar to the result shown in Fig.\ref{Fig6}(a),
Fig.\ref{Fig6}(b) shows that the absolute value of $R_{2}$
decreases quickly with the increment of symmetry breaking scale
$f$, and the absolute values of $R_{2}$ in the $LHT$ model can be
larger than $5\%$ in the range of $f<1.1~TeV$ for the three value
choices of $c_{\lambda}$ ($c_{\lambda}^{2} = 0.5$, $0.8$, $0.9$).
We can see from the figures that the most distinctive difference
between the relative deviation parameters $R_1$ and $R_2$, is that
the $LH$ result $R_1$ is always positive, while the $LHT$ result
$R_2$ is negative in our potted range of the symmetry breaking
scale $f$. That is due to the $t-\bar t-h^0$ coupling difference
between the $LHT$ and the $LH$ model.

\par
\begin{figure}[htb]
\centering
\includegraphics[scale=0.4]{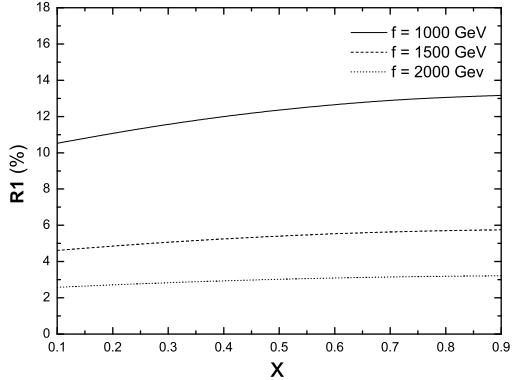}
\caption{ The relative deviation $R_{1}$ in the $LH$ model as the
functions of $x(\equiv4f\frac{v'}{v^2})$ with the conditions of
$c_{\lambda}^2 = 0.8$, $m_h=115~GeV$, $\sqrt{s}=800~GeV$ and three
value choices of the scale parameter $f$ (i.e., $f=1~TeV$,
$f=1.5~TeV$ and $f=2~TeV$) respectively.} \label{Fig6-1}
\end{figure}

In order to show the dependence of the relative deviation $R_{1}$
on the parameter $x(\equiv 4f\frac{v'}{v^2})$ with the fixed
symmetry breaking scale $f$($f=1~TeV$, $1.5~TeV$ and $2~TeV$), we
plot Fig.\ref{Fig6-1} by taking $c_{\lambda}^2 = 0.8$,
$m_h=115~GeV$ and $\sqrt{s}=800~GeV$. From the figure we can see
that all the three curves go up slowly with the increment of
parameter $x$. It shows that the relative deviation $R_{1}$ in the
$LH$ model is not very sensitive to parameter $x$ quantitatively.
Since the T-parity forbids the generation of a nonzero vacuum
expectation value $v'$ for the triplet scalar field (i.e., $v'=0$
and then $x=\frac{4fv'}{v^2}=0$.), there is no relationship
between the $t \bar th^0$ Yukawa coupling and the parameter
$x$(see Eq.(2.6)). So the plot of $R_2$ versus $x$ is absent.
\vskip 1cm

\par
\begin{figure}[htb]
\centering
\includegraphics[scale=0.35]{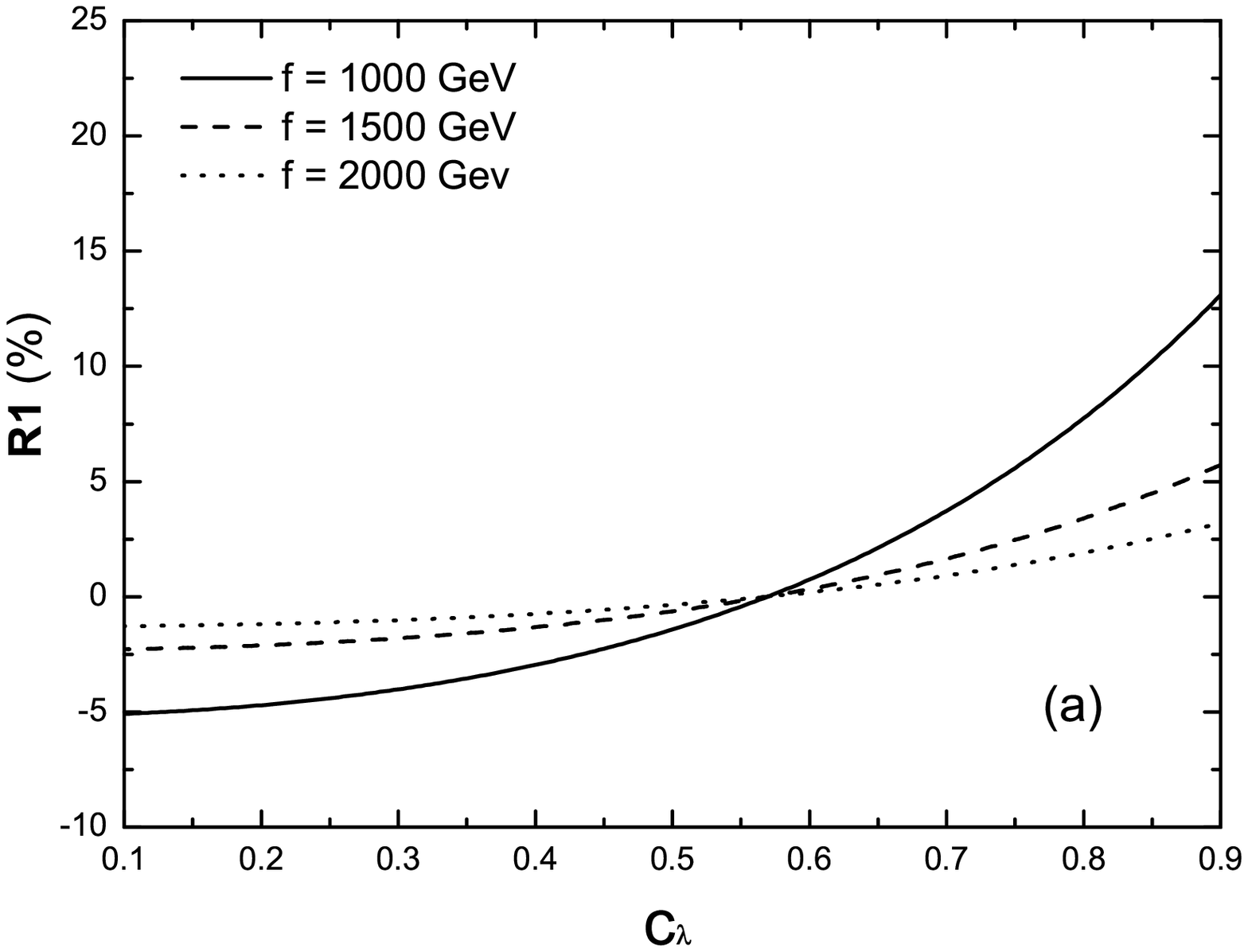}
\includegraphics[scale=0.35]{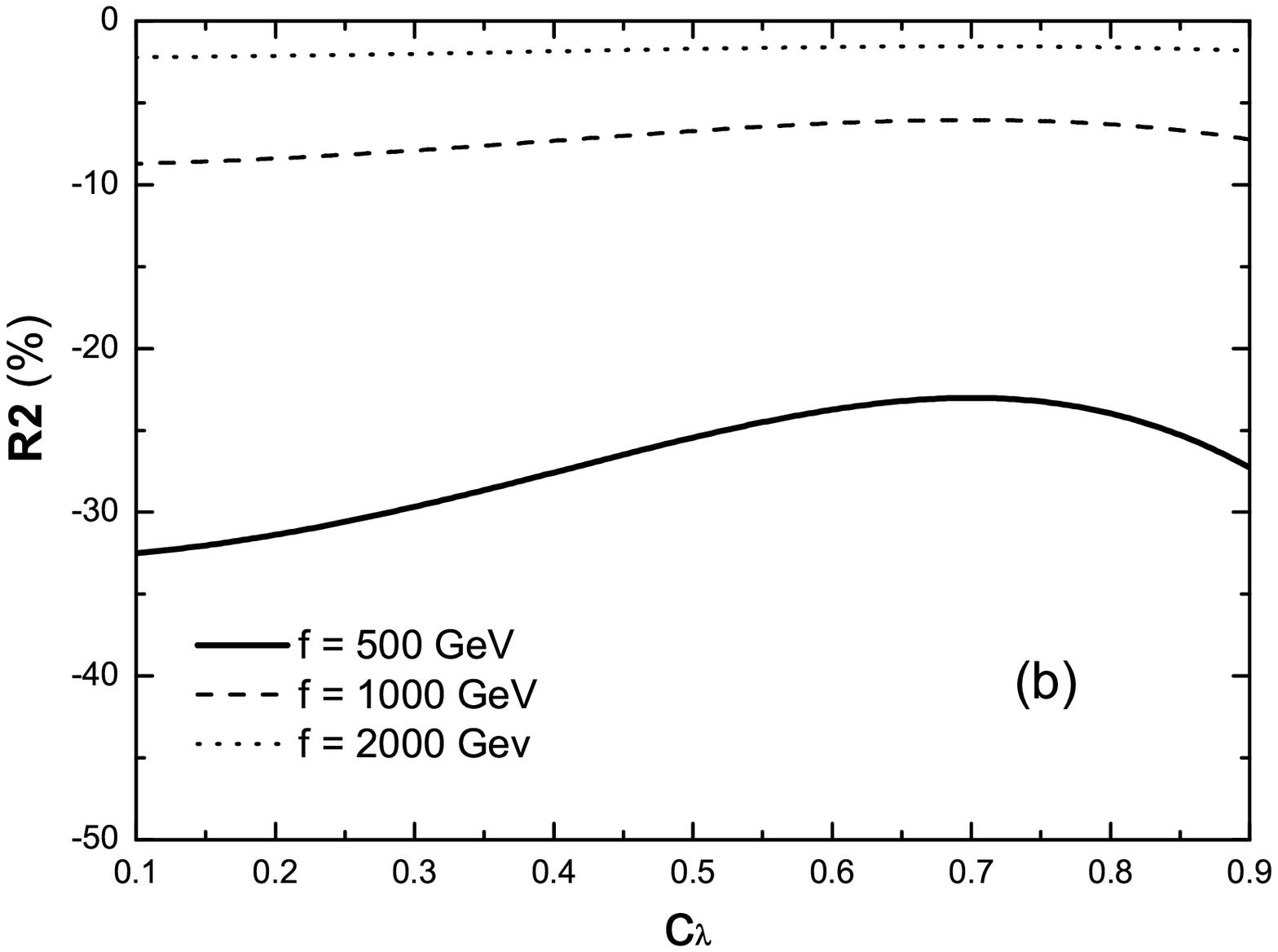}
\caption{ The relative deviations $R_{1}$ and $R_{2}$ as the
functions of parameter $c_{\lambda}$ for three value choices of
the symmetry breaking scale $f$ with $m_h=115~GeV$ and
$\sqrt{s}=800~GeV$. (a) is for $R_{1}$ in the $LH$ model, and (b)
is for $R_{2}$ in the $LHT$ model.} \label{Fig8}
\end{figure}

To study the dependence of the cross section for \eetth process on
the mixing parameter $c_{\lambda}$, we present the relative
deviation $R_{1}$ in the $LH$ model as a function of $c_{\lambda}$
in Fig.\ref{Fig8}(a), with $m_h=115~ GeV$, $\sqrt{s}=800~GeV$ and
$f=1~TeV$, $1.5~TeV$, $2~TeV$, respectively. One can read out from
the figure that the value of the relative deviation parameter
varies in a range from $-5\%$ to $13\%$ for $f=1~TeV$. And there
exists a special point of $c_{\lambda} = 0.57$, where the values
of $R_{1}$ for all the three choices of symmetry breaking scale
$f$ become zero. That is because with $c_{\lambda} = 0.57$, the
$t-\bar t-h^0$ coupling in $LH$ model converts into $SM$ one.
Moreover, for $c_{\lambda} < 0.57$ the values of the relative
deviation $R_{1}$ for $f=1~TeV$, $1.5~TeV$, $2~TeV$ are negative,
while they are positive when $c_{\lambda}
> 0.57$. In Fig.\ref{Fig8}(b), the relative deviation
$R_{2}$ generated by the $LHT$ model, is depicted as a function of
the mixing parameter $c_{\lambda}$ for three value choices of the
symmetry breaking scale $f$ (i.e. $f = 500~GeV$, $1~TeV$ and
$2~TeV$) with $m_h=115~GeV$ and $\sqrt{s}=800~GeV$. One can see
from Fig.\ref{Fig8}(b) that, when $f=500~GeV$, the absolute value
of $R_{2}$ can be beyond $30\%$ which might be easily observed at
the future $ILC$.

\par
\begin{figure}[htb]
\centering
\includegraphics[scale=0.35]{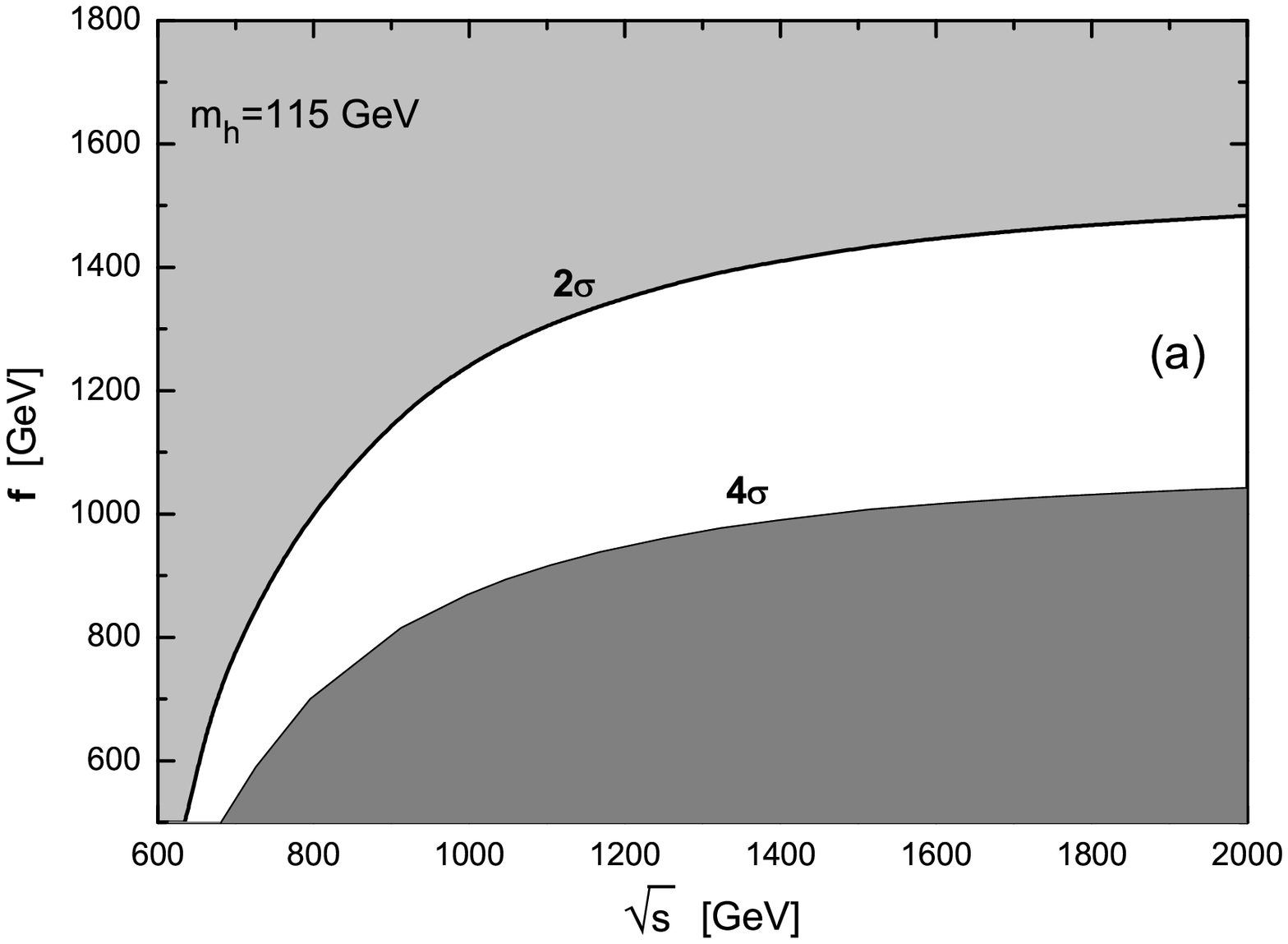}
\includegraphics[scale=0.35]{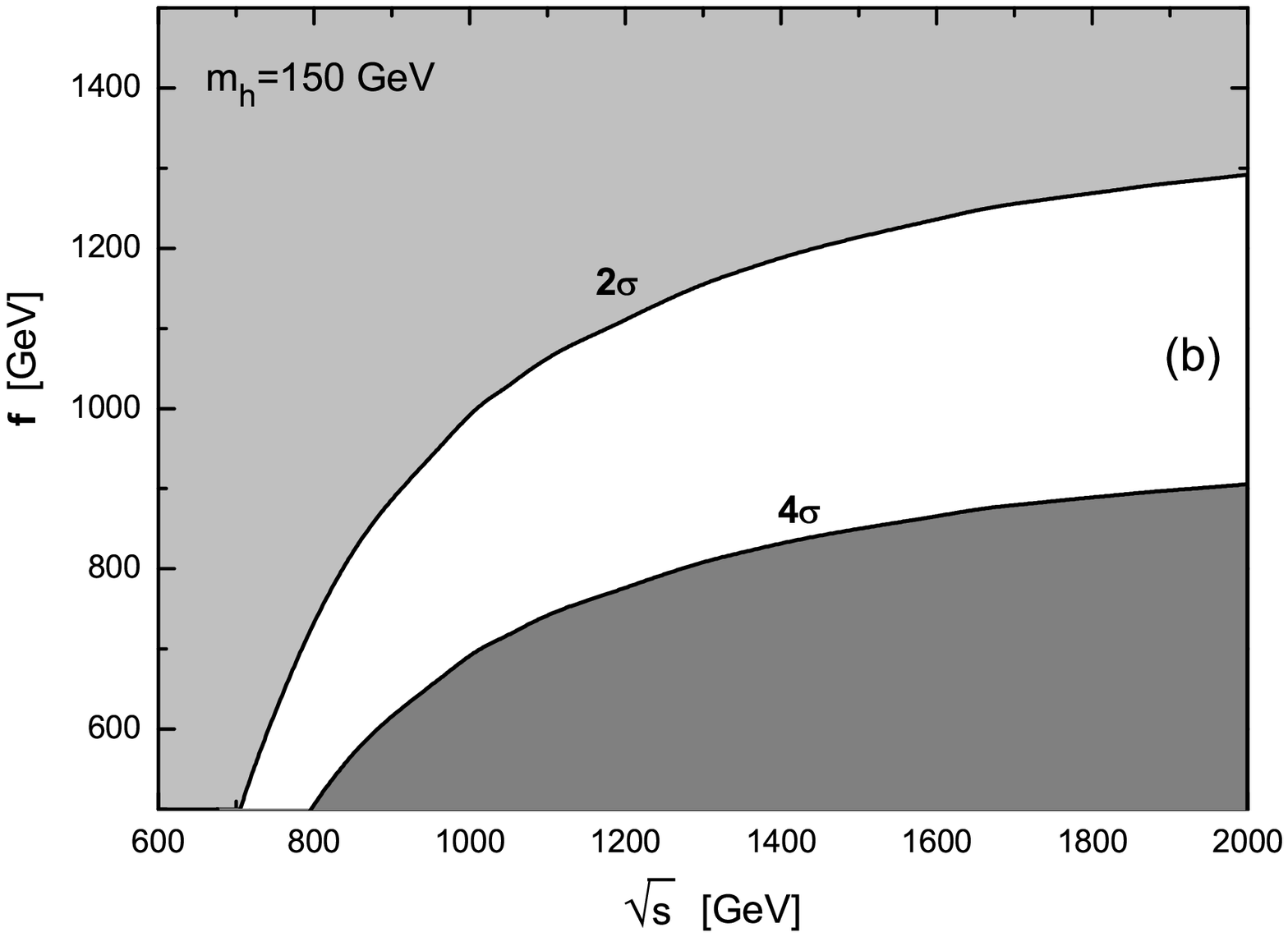}
\includegraphics[scale=0.35]{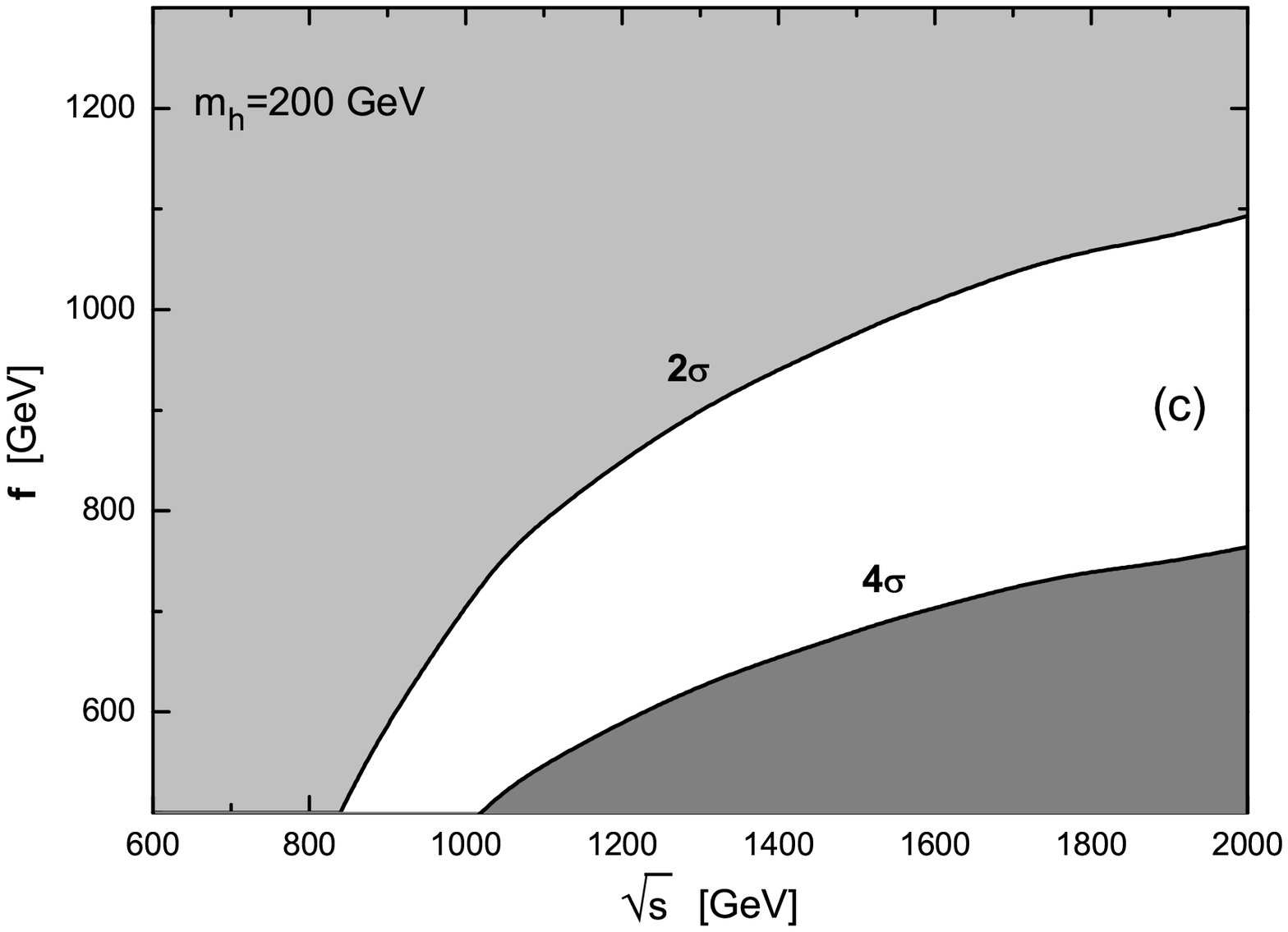}
\caption{ The $LH$ effect observation area(gray) and the $LH$
effect exclusion area (light gray) for the \eetth process in the
$\sqrt{s}-f$ parameter space with $c_{\lambda}^2 = 0.8$ and ${\cal
L}_{e^+e^-}=1000~fb^{-1}$. (a) is for $m_h=115~GeV$, (b) is for
$m_h=150~GeV$ and (c) is for $m_h=200~GeV$.}  \label{Fig9}
\end{figure}
\begin{figure}[htb]
\centering
\includegraphics[scale=0.35]{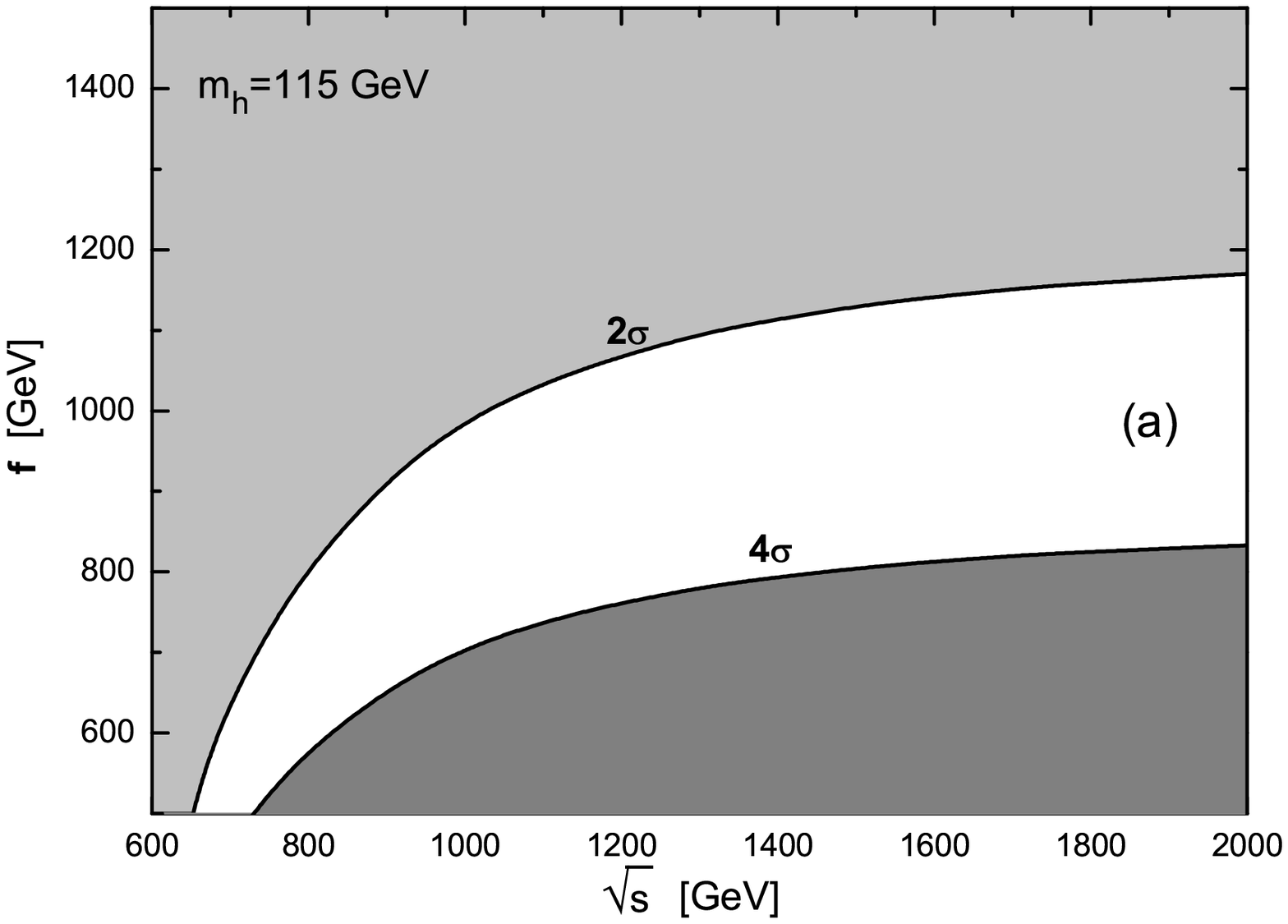}
\includegraphics[scale=0.35]{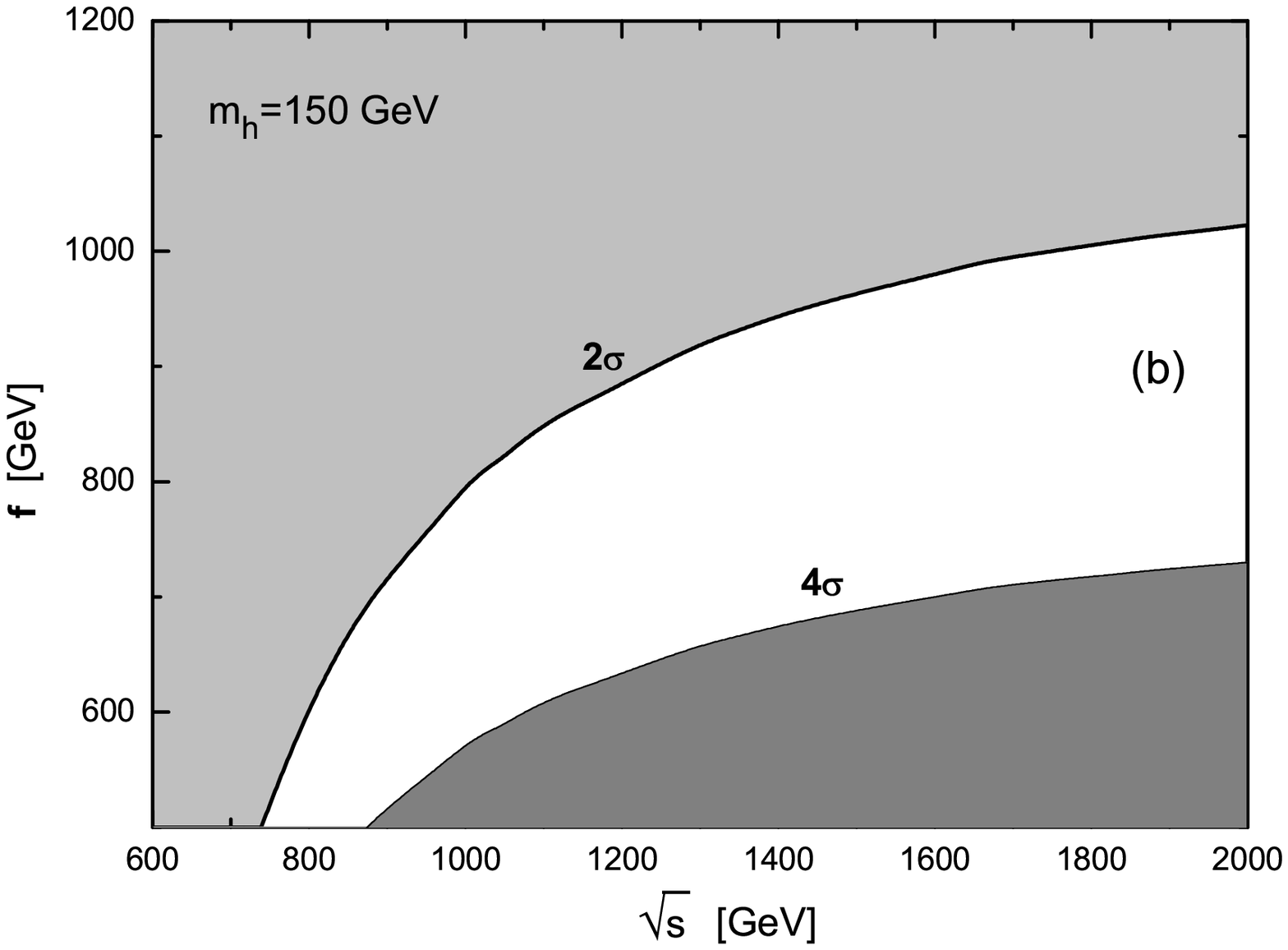}
\includegraphics[scale=0.35]{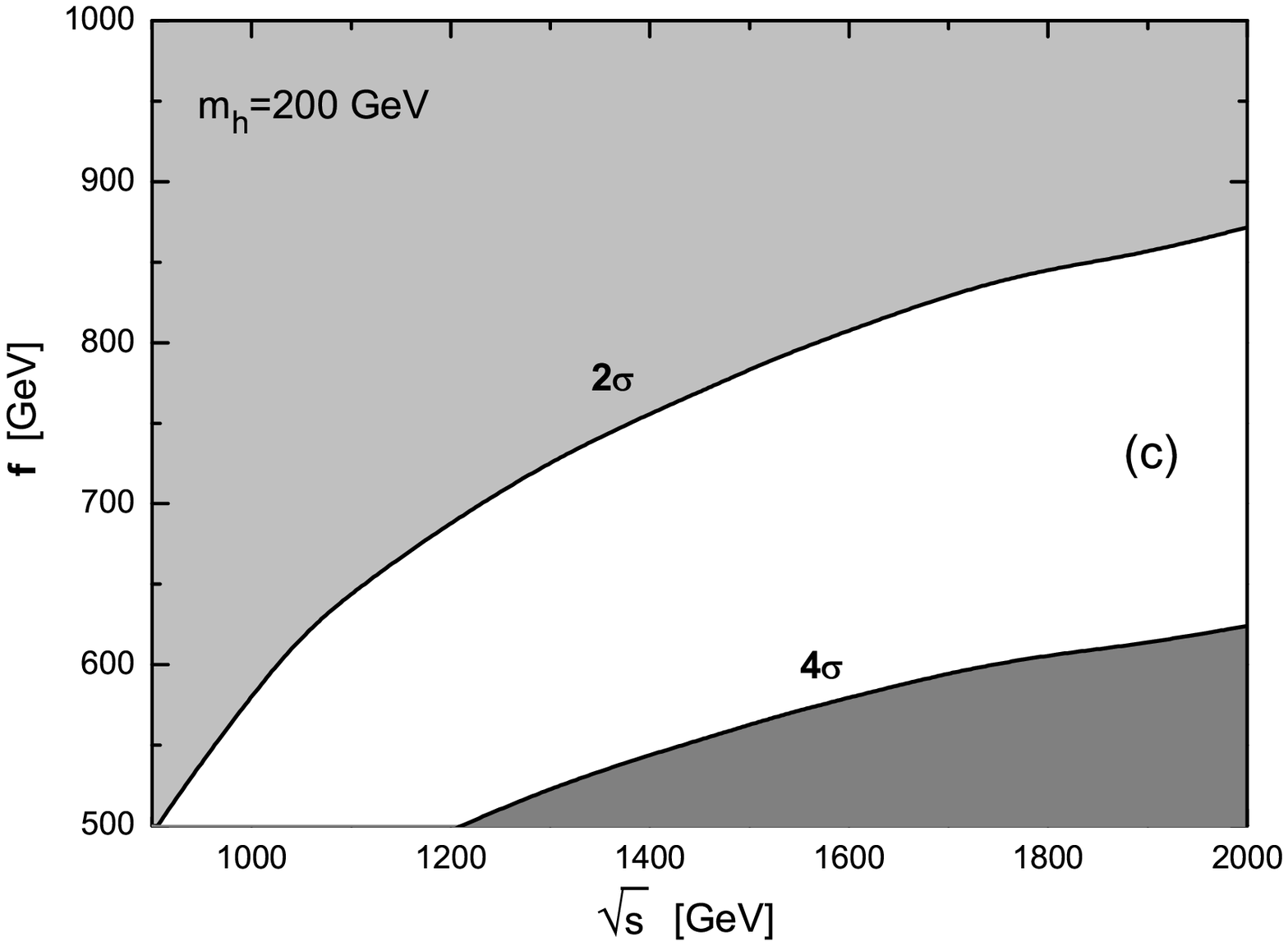}
\caption{ The $LHT$ effect observation area(gray) and the $LHT$
effect exclusion area (light gray) for the \eetth process in the
$\sqrt{s}-f$ parameter space with $c_{\lambda}^2 = 0.8$ and ${\cal
L}_{e^+e^-}=1000~fb^{-1}$. (a) is for $m_h=115~GeV$, (b) is for
$m_h=150~GeV$ and (c) is for $m_h=200~GeV$.}  \label{Fig10}
\end{figure}

As demonstrated in the above figures, both the $LH$ and $LHT$
models can obviously modify the cross section of the \eetth
process from the $SM$ prediction in some specific parameter
regions, if the $LH$ or $LHT$ really exists. Since the signals of
the $LH$ or $LHT$ model can be found only when the deviation of
the cross section from its $SM$ prediction, $\Delta
\sigma_{LH,LHT}$, is large enough, we assume that the $LH/LHT$
model effect can and can not be observed, only if
\begin{eqnarray}
\label{upper} \Delta\sigma_{LH,LHT}
=\sigma_{LH,LHT}^{NLO}-\sigma_{SM}^{NLO} \geq
\frac{4\sqrt{\sigma_{LH,LHT}^{NLO}{\cal L}}}{\cal L},
\end{eqnarray}
and
\begin{eqnarray}
\label{lower} \Delta\sigma_{LH,LHT}
=\sigma_{LH,LHT}^{NLO}-\sigma_{SM}^{NLO} \leq
\frac{2\sqrt{\sigma_{LH,LHT}^{NLO}{\cal L}}}{\cal L},
\end{eqnarray}
respectively. In the following discussions, we assume the $ILC$
integrated luminosity ${\cal L}_{e^+e^-}=1000~fb^{-1}$. We depict
the regions in the $\sqrt{s}-f$ parameter space in Fig.\ref{Fig9},
where the $LH$ effect can and cannot be observed from process
\eetth according to the above criteria, correspondingly. Figures
\ref{Fig9}(a), (b) and (c) correspond to taking $c_{\lambda}^2 =
0.8$, $m_h=115$, $150$ and $200~GeV$ respectively. In order to
show the deviation of the cross section in the $LHT$ model from
the $SM$ prediction, we also depict the regions in the
$\sqrt{s}-f$ parameter space in Figures \ref{Fig10}(a-c) by
adopting the same criteria used in Fig.\ref{Fig9}, with
$m_h=115~GeV$, $150~GeV$ and $200~GeV$ separately. In this figure,
the other input parameters are taken to be the same values as
discussed for Fig.\ref{Fig9}. Comparing Fig.\ref{Fig9} and
Fig.\ref{Fig10}, we can see clearly the difference of the effects
from the $LHT$ and $LH$ models. In Table 1 we list some typical
exclusion limits and corresponding $4\sigma$ observation limits on
$f$ and $\sqrt{s}$ according to the criteria shown in
Eqs.(\ref{upper}-\ref{lower}) for the \eetth process in the
$LH/LHT$ model, where most of the data for the $LH$ and $LHT$
model can be read out from Figs.\ref{Fig9}(a-c) and
Figs.\ref{Fig10}(a-c).

\par
\begin{figure}[htb]
\centering
\includegraphics[scale=0.35]{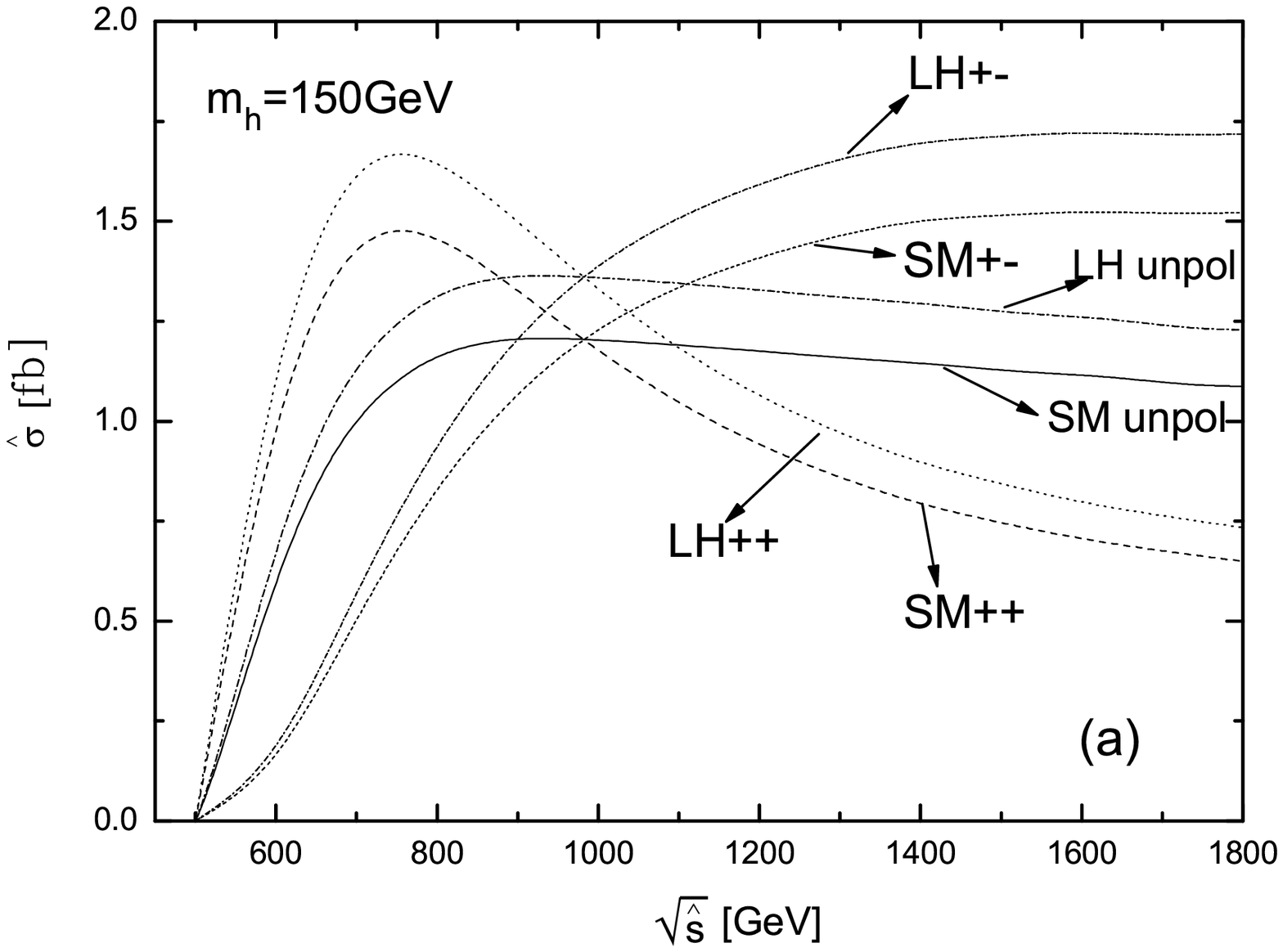}
\includegraphics[scale=0.35]{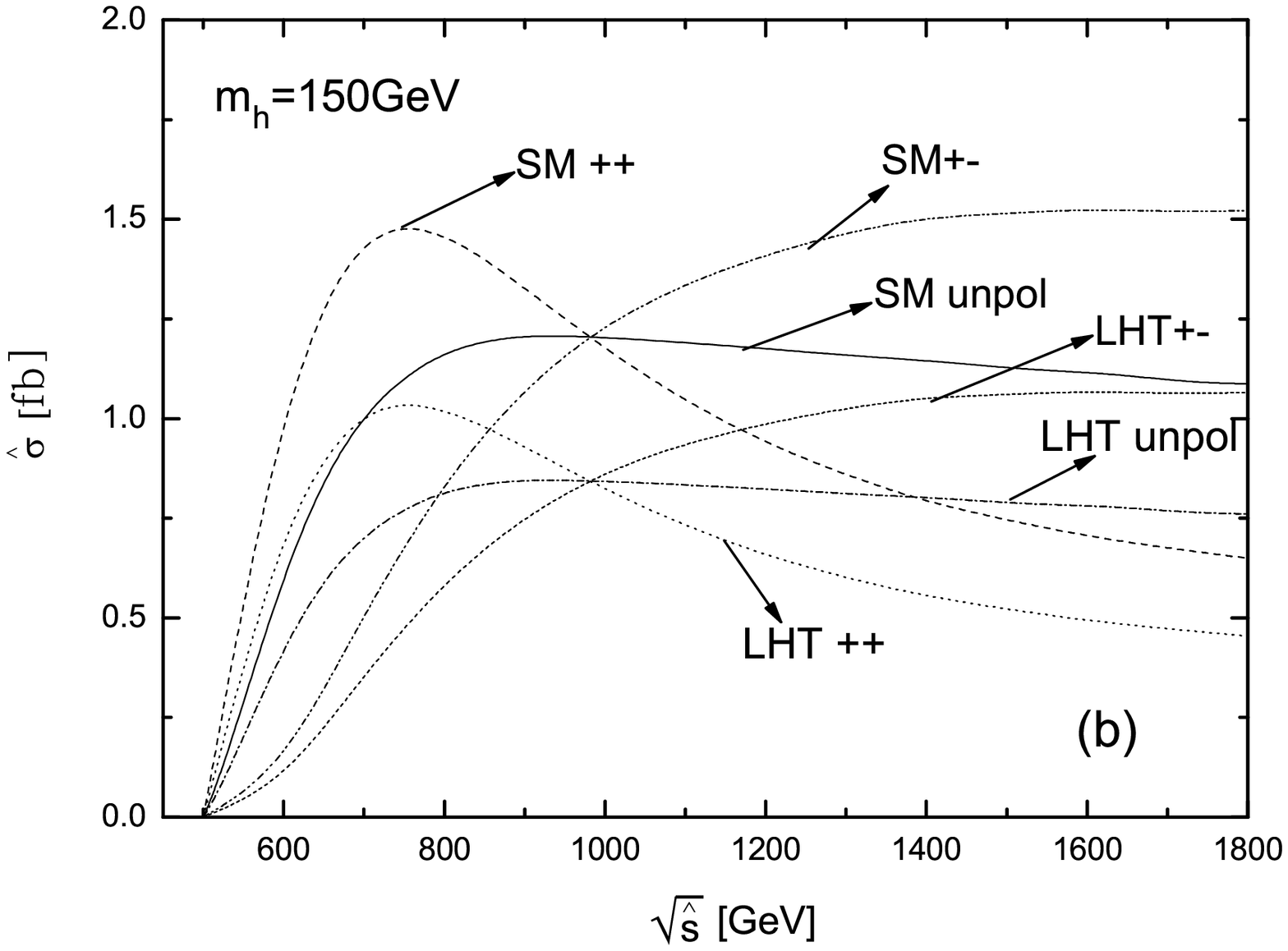}
\caption{ The cross sections of the process \rrtth as the
functions of $\gamma\gamma$ c.m.s. energy $\sqrt{\hat{s}}$ when
$m_h=150~GeV$ and $c_{\lambda}^2=0.8$. (a) is for the $LH$ model
with $f=1~TeV$, and (b) is for the $LHT$ model with  and
$f=0.5~TeV$.} \label{Fig11}
\end{figure}

In order to compare the production rates in different polarization
cases of initial photons for process \rrtth, we depict their cross
sections of process \rrtth as the functions of the $\gamma\gamma$
colliding energy $\sqrt{\hat{s}}$ in Fig.\ref{Fig11}(a) and (b) in
the frameworks of $LH$ and $LHT$ model separately, where
$m_h=150~GeV$, $f=1~TeV$ for the $LH$ model and $f=0.5~TeV$ for
the $LHT$ model, and the notation of $+~-$ represents helicities
of the two initial photons being $\lambda_1=1$ and $\lambda_2=-1$.
Since the cross-sections of the $+~-$ and $-~+$ photon
polarization (J=2) are equal, and also the cross-sections of the
$+~+$ and $-~-$ photon polarization (J=0) are the same, we only
present the total cross-sections in three cases in
Fig.\ref{Fig11}(a,b): $+~-$, $+~+$ and unpolarized photons. We can
see from the figures that the $LH/LHT$ model effects in the $+~+$
photon polarization case are obviously enhanced in comparison with
the unpolarized photon case in the vicinity of $\sqrt{\hat{s}}\sim
700~GeV$, while the $LH/LHT$ model effects in the $+~-$ case are
more significant when $1.2~TeV<\sqrt{\hat{s}}<1.8~TeV$.

\par
\begin{figure}[htb]
\centering
\includegraphics[scale=0.35]{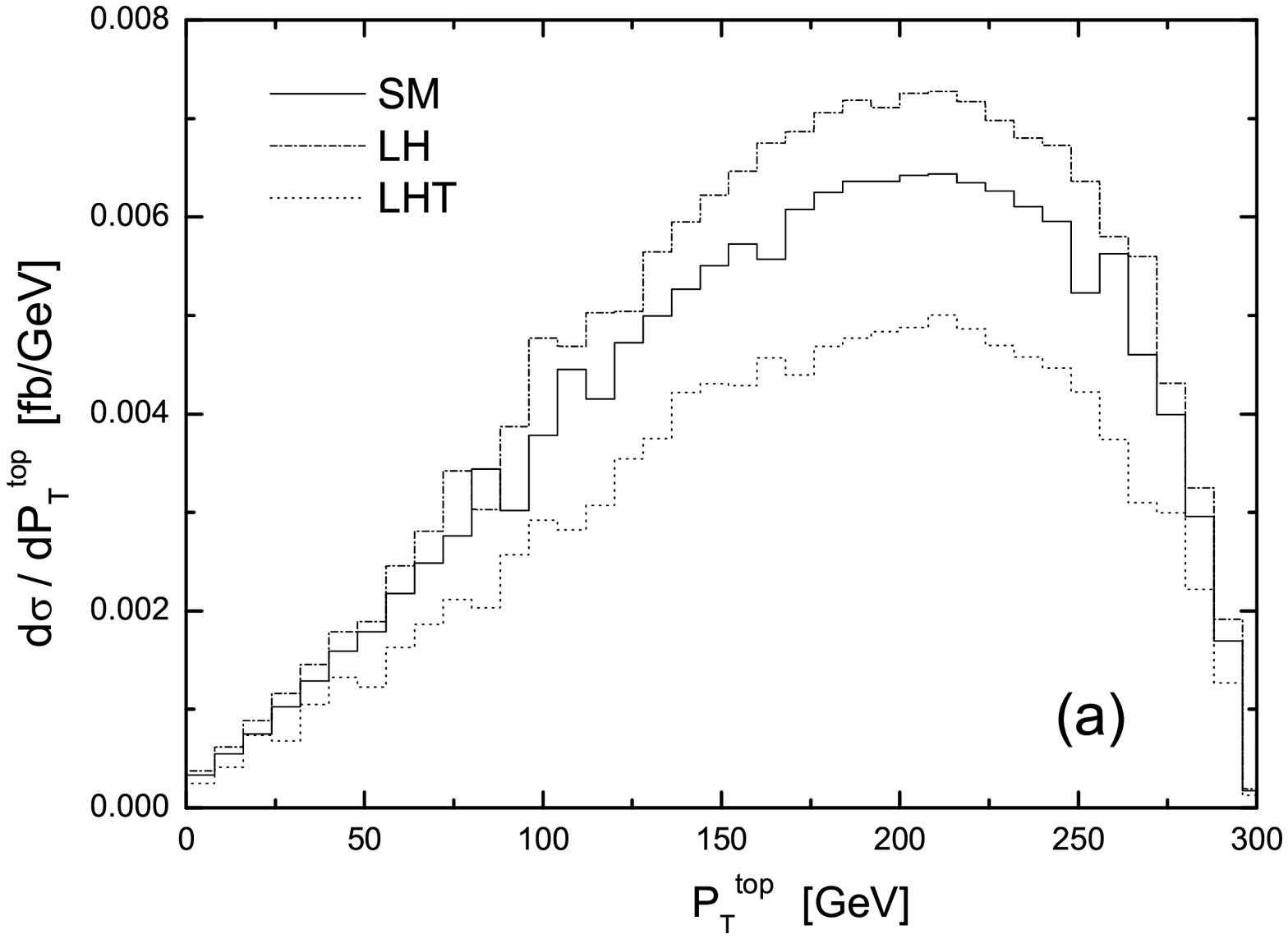}
\includegraphics[scale=0.35]{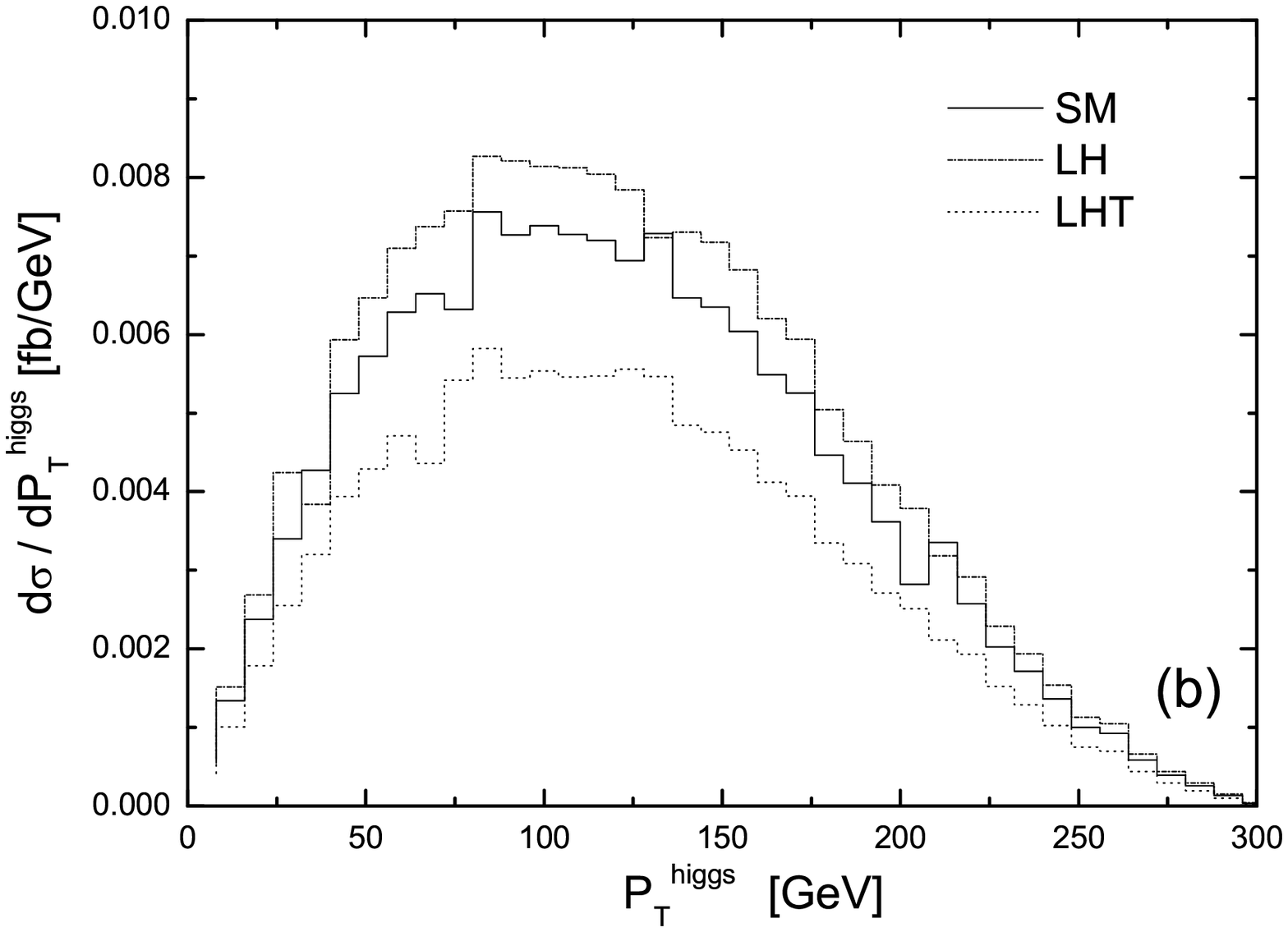}
\caption{ The transverse momentum distributions of the final state
particles($t,h^0$) at QCD NLO for the process \rrtth with
$c_{\lambda}^2 = 0.8$, $m_{h}=150~GeV$, $\sqrt{\hat{s}}=800~GeV$
and $f=1(0.5)~TeV$(in the $LH(LHT)$ model) at the ILC. (a) is for
the distributions of $p_T^t$, and (b) is for the distributions of
$p_T^h$. } \label{Fig12}
\end{figure}

In Fig.\ref{Fig12}(a-b), we plot the distributions of the
transverse momenta of the final states($p_T^t$ and $p_T^h$) for
the process \rrtth with $c_{\lambda}^2 = 0.8$, $m_{h}=150~GeV$,
$\sqrt{\hat{s}}=800~GeV$ and $f=1(0.5)~TeV$(in the $LH(LHT)$
model) at the ILC. Due to the CP-conservation, the distributions
of the transverse momentum of anti-top quark, $p_T^{\bar t}$, in
the process \rrtth should be the same as that of
$d\sigma_{SM}/dp_T^t$ shown in Fig\ref{Fig12}(a). These figures
demonstrate that the $LH$ and $LHT$ model corrections
significantly modify the $SM$ distributions of the differential
cross sections $d\sigma_{SM}/dp_T^t$ and $d\sigma_{SM}/dp_T^h$ at
the ILC, respectively. We find that in the regions around $p_T^t
\sim 200~GeV$ and $p_T^h\sim 100~GeV$, the $LH/LHT$ corrections
can be more significant than in other regions.

\begin{table}[htbp]
\begin{center}
\begin{tabular}{c|c|c|c}
\cline{1-4}
$\sqrt{s}$&~~~~$m_h$~~~&~~$LH$:~~$f$~[GeV]~&~~~~$LHT$:~~$f$~[GeV]~\\[0mm]
\cline{3-4}\vspace*{-0ex}[TeV]&~~~[GeV]~~~&~~ $2\sigma$,~~~~~$4\sigma$~~&~~ $2\sigma$,~~~~~$4\sigma$ \\
\cline{1-3} \hline
~~& ~~115~~&1004,~~~~~712~&~~803,~~~~578 \\[0mm]
\cline{2-4}
0.8~~& ~~150~~&~740,~~~~~500~&~~602,~~~~466 \\[0mm]
\cline{2-4}
~~& ~~200~~&~467,~~~~~390~&~~500,~~~~446 \\[0mm]
\cline{1-3} \hline
~~& ~~115~~&1246,~~~~~875~&~~985,~~~~708 \\[0mm]
\cline{2-4}
1.0~~& ~~150~~&~991,~~~~~688~&~~796,~~~~570 \\[0mm]
\cline{2-4}
~~& ~~200~~&~707,~~~~~488~&~~592,~~~~462 \\[0mm]
\cline{1-3} \hline
~~& ~~115~~&1429,~~~~1009~&~1130,~~~~810 \\[0mm]
\cline{2-4}
1.5~~& ~~150~~&1216,~~~~~~851~&~~966,~~~~692 \\[0mm]
\cline{2-4}
~~& ~~200~~&~978,~~~~~~683~&~~784,~~~~566 \\[0mm]
\hline
\end{tabular}
\caption{ The dependence of the $2\sigma$ exclusion limits and
corresponding $4\sigma$ observation limits on $f$ and $\sqrt{s}$
for the \eetth process with $c_{\lambda}^2 = 0.8$ in the $LH$ and
$LHT$ model. }
\end{center}
\end{table}

\par
\section{Summary}
\par
We investigated the effects of the littlest models with and
without T-parity including the QCD NLO corrections, on the
associated $t\bar th^0$ production process \eetth at future
electron-positron linear colliders. We present the regions of
$\sqrt{s}-f$ parameter space in which the $LH$ and $LHT$ effects
can and cannot be discovered with the criteria assumed in Eqs.
(\ref{upper}) and (\ref{lower}). The production rates of process
\rrtth in different incoming photon polarization collision modes
are also discussed. We find that the measurement of the process
\rrtth in polarized photon collision mode is of benefit to
discovering the effects of the $LH/LHT$ model in some specific
c.m.s. energy ranges. We discover that the effects of the $LHT$
model in the process \eetth generally can be greater than in the
$LH$ model when the symmetry breaking scale $f$ has a relative
small value due to the $t-\bar t-h^0$ coupling difference between
the $SM$, $LHT$ and the $LH$ model. Our results show that the
relative deviation $R_1$ for the $LH$ model in the process \eetth
is always positive, while $R_2$ for the $LHT$ model is negative in
our chosen range of the symmetry breaking scale $f$. We conclude
that the future experiment at the $ILC$ could discover the effects
on the \eetth cross section contributed by the $LH$ or $LHT$ model
in some parameter space, or put more stringent constraints on the
$LH$/$LHT$ parameters.

\vskip 5mm
\par
\noindent{\large\bf Acknowledgments:} This work was supported in
part by the National Natural Science Foundation of China, the
Education Ministry of China and a special fund sponsored by
Chinese Academy of Sciences.

\end{document}